\documentclass[twocolumn,trackchanges,twocolappendix]{aastex701}

\newcommand{\xmm}{\textit{XMM-Newton}}
\newcommand{\epn}{EPIC-pn}

\newcommand{\rgs}{RGS}
\newcommand{\swift}{\textit{Swift}}
\newcommand{\xrism}{\textit{XRISM}}
\newcommand{\nustar}{\textit{NuSTAR}}

\newcommand{\fekalpha}{Fe\,K$\alpha$}

\hypersetup{linkcolor=blue,citecolor=blue,filecolor=blue,urlcolor=blue}

\shorttitle{Disk-Corona Connection and Broad Fe\,K$\alpha$ Variability in Mrk 766}
\shortauthors{A. Mondal et al.}

\received{May 4, 2026}
\revised{August 25, 2026}
\accepted{September 14, 2026}

\begin{document}

\title{Dynamic Accretion Disk-Corona Connection and Broad Fe\,K$\alpha$ Variability in Mrk 766 Revealed by Time-Resolved High Resolution X-Ray Spectroscopic Analysis}

\author[orcid=0009-0007-8215-6031,sname=Mondal,gname=Ayon]{Ayon Mondal}
\affiliation{Department of Physics, University of Maryland Baltimore County, 1000 Hilltop Circle, Baltimore, MD 21250, USA.}
\affiliation{School of Astrophysics, Presidency University, 86/1 College Street, Kolkata-700073, India.}
\email[show]{ayonm1@umbc.edu}

\author[orcid=0000-0001-5647-3366,sname=Saha,gname=Tathagata]{Tathagata Saha}
\affiliation{Inter-University Centre for Astronomy and Astrophysics (IUCAA), PB No.4, Ganeshkhind, Pune-411007, India.}
\email[show]{tathagata.saha@iucaa.in}

\author[orcid=0009-0005-6387-4789,sname=Sar,gname=Arijit]{Arijit Sar}
\affiliation{School of Astrophysics, Presidency University, 86/1 College Street, Kolkata-700073, India.}
\email[]{arijit.rs@presiuniv.ac.in}

\author[orcid=0000-0001-9899-7686,sname=Chatterjee,gname=Ritaban]{Ritaban Chatterjee}
\affiliation{School of Astrophysics, Presidency University, 86/1 College Street, Kolkata-700073, India.}
\email[]{ritaban.astro@presiuniv.ac.in}

\correspondingauthor{Tathagata Saha}
\email{tathagata.saha@iucaa.in}

\begin{abstract}

    We present a multi-epoch, primarily X-ray study of the narrow-line Seyfert 1 (NLSy1) galaxy Mrk~766 by combining archival \textit{XMM-Newton} observations spanning nearly two decades together with a \textit{XRISM} observation obtained during its Performance Verification Phase. We analyzed the absorption and emission features in the soft X-ray spectra from RGS and EPIC-pn data, revealing a stratified multi-phase potentially ionised absorber with a column density varying at a timescale of $\sim$1 day. The covering fraction shows positive correlation with the continuum flux, however with significant intrinsic scatter, between the epochs. Mrk~766 exhibits a multi-component Fe\,K$\alpha$ emission line as revealed by fits to the \textit{XRISM} data. A quasi-phenomenological model comprising a torus and a broad \textit{diskline} component distinguishes the two components and constrains the inner radius of the potential broad line emitting disk substructure between $R_{in}=40-60\,r_{\rm g}$. Analysis of multiple epochs of the Fe K$\alpha$ complex across \textit{XMM-Newton} observations reveals that the broad component tracks the continuum flux closely and effectively reflects the trend with respect to the continuum. The narrow component is consistent with a distant emitter, potentially the torus. Quasi-simultaneous monitoring with \textit{Swift}-XRT and UVOT further indicates UV emission lagging the X-rays by $5.9^{+4.1}_{-5.7}$ hours, consistent with disk reprocessing of coronal X-ray fluctuations, and corresponds to a light travel distance of up to $10^{3}\,r_{\rm g}$. Overall, our analysis maps the extent of the dynamically evolving absorber, the iron line emitter, and the accretion disk in this highly accreting NLSy1 system.

\end{abstract}



\section{Introduction}
\label{sec:intro}

    The X-ray spectra of Seyfert 1 AGNs commonly exhibit a prominent \fekalpha\ line complex, originating from reprocessing of the primary X-ray continuum in the circumnuclear environment surrounding the central supermassive black hole \citep{Costantini_2007MNRAS, Mehdipour_2010A&A}. Observations with \xmm\ have revealed that this complex often consists of a narrow core \citep[e.g.][]{Murphy_Yaqoob_2009, Yaqoob_2012}, accompanied by excess broad emission feature that is interpreted as relativistically broadened reflection originating in the inner accretion disk \citep[e.g.,][]{Nandra_2007, Bhayani_Nandra_2011}. In addition to these reflection features, Seyfert 1 spectra frequently display signatures of ionized absorption in the soft X-ray band ($<2 \,\mathrm{keV}$), indicating the presence of multi-phase, outflowing material along the line of sight \citep{Costantini_2007MNRAS, Mehdipour_2010A&A}. Taken together, these components make the \fekalpha\ complex a powerful diagnostic of AGN reprocessing, with the potential to probe spatially distinct regions of the circumnuclear environment around the black hole ranging from the inner accretion disk, broad-line region (BLR) to more distant structures such as the torus and large-scale outflows \citep[e.g.,][]{George_Fabian_1991, Yaqoob_Padmanabhan_2004, Hagai_2015}. In Narrow-Line Seyfert 1 (NLSy1) galaxies, this picture is more complicated since they exhibit rapid continuum variability and strong ionized absorption, which can significantly affect the observed \fekalpha\ emission. The rapid variability points to a compact, rapidly evolving corona and consequently a variable inner-disk reflection \citep{J_Miller_2007, Fabian_Ross_2010, Uttley_2014}. Disentangling these various contributions remains challenging, however, due to degeneracies inherent in spectral modeling as well as limitations imposed by current instrumental capabilities.

    The moderate spectral resolution of CCD detectors blends multiple narrow components of the \fekalpha\ line, limiting our ability to resolve its intrinsic structure. Consequently, the relative contributions of inner-disk reflection, distant reprocessing, and clumpy ionized absorbers in shaping the observed \fekalpha\ complex remained unexplored. The advent of high-resolution X-ray spectroscopy with the \xrism-Resolve\ micro-calorimeter provides a significant advance in this context, delivering an unprecedented spectral resolution of $\Delta E \simeq 4.5\text{--}4.9~\rm{eV}$ (FWHM) at 6 keV \citep{Tashiro_2021, Ishisaki_2022, Kelley_2025_Resolve}. This capability enables the decomposition of the \fekalpha\ line into distinct kinematic components, offering the potential to map the geometry of the reprocessing regions that produces the \fekalpha\ line complex, and to constrain the physical extent of each emission region, e.g., the disk, BLR or torus.

    Mrk 766, a nearby Narrow-Line Seyfert 1 (NLSy1) AGN ($z \sim 0.0129$; \citealt{Falco_1999PASP}) hosted in a barred spiral galaxy with a SMBH mass of $\rm M_{\bullet} = 1.26^{+1.00}_{-0.77} \times 10^6\, M_\odot$ \citep{Bentz_2009ApJ, Bentz_2010ApJ, Giacche_2014A&A}, has been a focus of multiple X-ray studies owing to its spectral and flux variability across a broad energy range. Observations with \xmm, \nustar, and \swift\ have revealed signatures of both ionised absorption and relativistic reflection, along with variability driven by partial covering by clumpy absorbers \citep{Risaliti_2009ApJ, Risaliti_2009MNRAS}. The source's soft X-ray spectrum shows evidence for multi-phase warm absorbers, and potentially even collisionally ionized gas, as well as emission lines that may originate from extended photoionized regions or be linked to the same outflows. Debate has persisted regarding the interpretation of spectral features near the Fe~K band (6--7~keV). Evidence for ionized Fe absorption lines was first presented by \citet{Pounds_2003MNRAS}, and subsequent studies have firmly established the presence of blue-shifted, ionized FeK absorption, an indicator of outflowing ionized gas along our line of sight (e.g., \citealt{Miller_2007, Turner_2007, Risaliti_2011, Liebmann_2014ApJ, Buisson_2018}). Some studies attribute these to relativistically-broadened reflection from the inner accretion disc \citep{Branduardi-Raymont_2001A&A}, while others favor a hybrid model which includes both partial-covering absorption and reflection \citep[e.g.][]{Buisson_2018, Mochizuki_2023, Zatarain_2025}. A stronger indication of relativistic reflection in Mrk~766 arises from the detection of time delay between the primary continuum and the reflected emission interpreted as caused by the light travel time between the corona and the accretion disc \citep{Emmanoulopoulos_2011,DeMarco_2013, Kara_2016}. For Mrk 766, quasi-continuous XMM-Newton monitoring showed that spectral variability is driven by broad-line region clouds transiting the line of sight, implying the presence of associated highly ionized gas and an outflow component \citep{Risaliti_2011}.
    
    Mrk~766 has been observed with \xmm's RGS and EPIC-pn, revealing both warm absorbers and possible reflection or partial covering effects. However, the relationship between variability in partial covering clouds and warm absorbers, the broad and narrow \fekalpha\ emission components, and the intrinsic X-ray continuum flux remains poorly constrained.
    
    In this study, we present a joint \xrism\ and \xmm\ analysis of Mrk 766 using archival \xmm\ data from the RGS and EPIC-pn detectors and a \xrism\ observation during the Performance Verification Phase, with particular focus on a detailed study of the \fekalpha\ complex and identifying and characterizing the ionized absorbers. We connect the \fekalpha\ emission to the properties of the continuum and wind and understand how the morphology of the line-emitting region changes with the accretion rate and outflow properties. While Mrk~766 has been the subject of numerous spectral studies, little attention has been given to its temporal behavior in the X-ray and UV bands. Simultaneous monitoring with \swift–XRT and UVOT helps us to study correlated variability between the accretion disk and corona.

    This paper is organized as follows. Observation and data reduction methods are described in Section~\ref{sec:data_reduc}. The multi-epoch X-ray spectra analysis is described in Section~\ref{sec:x-ray_spec}. The results are presented in Section~\ref{sec:results}, the implications are discussed in Section~\ref{sec:discussion}, and the conclusions are summarized in Section~\ref{sec:conclusions}.

    Throughout this work, we adopt the following $\Lambda\rm{CDM}$ cosmology with: $H_0 = 70~\mathrm{km~s^{-1}~Mpc^{-1}}$, $\Omega_{\mathrm{m}} = 0.3$, and $\Omega_{\Lambda} = 0.7$ \citep{Planck_2020}. The analysis was carried out using the X-ray spectral fitting package \texttt{XSPEC v12.15.0} \citep{Arnaud_Xspec}. To evaluate the goodness of fit, we used the C-statistic \citep{Cash1979} and the $\chi^2$-statistic. To account for parameter degeneracies during the fitting process, we employed Markov Chain Monte Carlo (MCMC) techniques (Emcee; \citealt{emcee_2013}). The reported uncertainties were calculated at the 90\% confidence level unless otherwise stated.

\section{Observation and Data Reduction} \label{sec:data_reduc}

    Mrk~766 has been observed over the last two decades using multiple X-ray telescopes. In this paper, we focus primarily on X-ray observations obtained using \xmm\ between 2000 and 2015, and \xrism\ observation in 2024, along with intensive \swift\ monitoring during mid-2024. The datasets are listed in Table~\ref{tab:data_observations}.
    
\begin{deluxetable}{llcc}
\tablecaption{Summary of \xmm, \xrism, and \swift\ observations used in this work\label{tab:data_observations}}
\tablewidth{0pt}
\tabletypesize{\footnotesize}
\tablehead{
\colhead{Obs.} & \colhead{Obs. ID} & \colhead{Obs. Date} & \colhead{Exposure (s)}
}
\startdata
XMM1 & 0096020101 & 2000-05-20 & 58835 \\
XMM2 & 0109141301 & 2001-05-20 & 129906 \\
XMM3 & 0304030101 & 2005-05-23 & 95510 \\
XMM4 & 0304030301 & 2005-05-25 & 98910 \\
XMM5 & 0304030401 & 2005-05-27 & 98918 \\
XMM6 & 0304030501 & 2005-05-29 & 95514 \\
XMM7 & 0304030601 & 2005-05-31 & 98918 \\
XMM8 & 0304030701 & 2005-06-03 & 35017 \\
XMM9\tablenotemark{a} & 0763790401 & 2015-07-05 & 29300 \\
\hline
\xrism & 300007010 & 2024-06-24 & 111753 \\
\hline
\swift\tablenotemark{b} & 00030846043-- & 2024-05-10-- & \nodata \\
    & 00030846108 & 2024-06-08 & \\
\enddata
\tablenotetext{a}{Full-window mode observation}
\tablenotetext{b}{All \swift\ observations between 10 May -- 08 June 2024}
\end{deluxetable}

\subsection{\xmm}
\subsubsection{\epn}
\label{subsubsec:EPIC}

    Mrk~766 was observed multiple times with \xmm\ \citep{jansen_2001} between May 2000 and July 2015 (Table~\ref{tab:data_observations}). For all observations except XMM9, the European Photon Imaging Camera (EPIC-pn; \citealt{EPIC_2001}) was operated in the small-window mode; the XMM9 observation was performed in full-window mode.
    Data reduction was performed with the \xmm\ Science Analysis System (\texttt{SAS} v22.1.0; \citealt{gabriel_2004}) and \texttt{HEASoft v6.35}, using the most recent calibration files. 
    The raw observation data files were processed with the \texttt{epproc} and \texttt{emproc} tasks to generate calibrated event lists. 
    For all observations, conservative background cuts for flare removal retained a total GTI between 34\% to 66\% of the total exposure.
    Source spectra were extracted from circular regions with radii of $35$--$50\arcsec$, while background spectra were obtained from nearby source-free circular regions on the same CCD with typical radii between $80$ and $120\arcsec$. Standard flag filtering (\texttt{XMMEA\_EP}) was applied, and events were restricted to the $0.3$--$10$~keV range. Photon pile-up was checked with {\tt epatplot} and found to be negligible in all datasets.
    For EPIC-pn, spectra were extracted separately for single (pattern~0) and double (patterns~1-4) events.

\subsubsection{\rgs}
\label{subsubsec:RGS}

    We analyzed data from the Reflection Grating Spectrometers (RGS; \citealt{Herder_2001}) onboard \xmm, operating in the Spectroscopy High Event Rate (HER) and Small Event Size (SES) mode. The data reduction was performed using the Science Analysis System (SAS; version 22.1.0) with the latest calibration files available at the time of processing. The standard RGS pipeline processing was executed using the \texttt{rgsproc} task. To ensure data quality, we filtered background flaring periods by generating good-time intervals (GTIs) with the \texttt{tabgtigen} tool, applying a count rate threshold on the high-energy background. We extracted the first-order source and background spectra for both RGS1 and RGS2. Standard selection criteria were applied, including the recommended CCD patterns and spatial extraction regions. The response matrices were generated automatically by \texttt{rgsproc}. Following extraction, we combined the individual spectra and response files from RGS1 and RGS2 using the \texttt{rgscombine} task to improve signal-to-noise. This resulted in three primary files: the RGS1+RGS2 combined first-order source spectrum, background spectrum, and response matrix file for further analyses.

\subsection{\xrism}
\label{subsec:xrism_data_analysis}

    X-ray Imaging and Spectroscopy Mission (\xrism; \citealt{Xrism_science_paper, Tashiro_2025PASJ_XRISM}) observation of Mrk~766 (Obs ID: 300007010) in Performance Verification Phase (PVP) was carried out in 2024 using the Resolve \citep{Ishisaki_2022, Kelley_2025_Resolve} instrument with the filter wheel set to open position and simultaneously the Xtend \citep{Xtend, Noda_2025PASJ_Xtend} instrument collected data with CCD1 \& CCD2 operated in 1/8 window mode and CCD3 \& CCD4 were configured in full window mode. We followed the procedures described in the \xrism\ Quick Start Guide (v3.1)\footnote{\xrism\ QSG v3.1: \url{https://heasarc.gsfc.nasa.gov/docs/xrism/analysis/quickstart/index.html}} provided by \xrism\ Science Data Center for analysis and reduction of the raw data. The raw XRISM data were reprocessed using \texttt{xapipeline} separately for Resolve and Xtend, producing cleaned event files.
    Data analysis was performed using the latest available \xrism\ CALDB version-12 (20250915 release) and the ftools package consolidated in the HEASoft v.6.35 software package. Only high-resolution primary (Hp) events were retained for the Resolve spectral analysis. Of the 36 detector pixels in the Resolve instrument, Pixel 12 was excluded as it functions as the calibration pixel, and Pixel 27 was removed due to its unstable gain jumps. To eliminate contamination from pseudo low-resolution secondary (Ls) events and avoid response normalization errors, all Ls events were discarded before generating the RMF. The RMF for the Resolve spectrum was produced using the \texttt{rslmkrmf} task. A large full-array RMF (`L' configuration) was adopted, to provide an effective balance between spectral accuracy and computational efficiency.

    The Xtend cleaned event files were obtained from the standard \xrism\ \texttt{xapipeline}. Additional screening was applied to mitigate soft-band contamination from high-signal cosmic ray event and high particle background periods. Source and background spectra from the Xtend observations were extracted, the redistribution matrix file (RMF) for Xtend spectrum was generated with \texttt{xtdrmf}, and the ancillary response file (ARF) was produced with \texttt{xaarfgen} in point-source mode, using the exposure map generated via \texttt{xaexpmap}.

\subsection{Neil Gehrels Swift Observatory}
\label{subsec:Swift}
\subsubsection{\textit{Swift}-\rm{XRT}}
\label{subsubsec:XRT}

    Mrk~766 was monitored with the X-Ray Telescope (XRT; \citealt{Swift_XRT}) onboard \textit{Swift} simultaneously with the UVOT. We used all available photon counting (PC) mode data obtained during the month-long observing period of a total of 55 observations from 2024 May 10 to June 08 to construct the 0.3--10 keV light curve. The XRT light curve was generated using the online \swift-XRT data products generator tool\footnote{\swift-XRT data products generator tool: \url{https://www.swift.ac.uk/user_objects/}} provided by the UK \swift\ Science Data Center \citep{Evans_2009}.

\subsubsection{Swift-\rm{UVOT}}
\label{subsubsec:UVOT}

    Mrk~766 was observed with the Swift Ultraviolet/Optical Telescope (UVOT; \citealt{Roming_2005_UVOT}) simultaneously with the XRT. To construct the light curves in the U, UVW1, UVM2 and UVW2 bands, we utilized all available archival Swift observations over a month-long period. Data reduction was carried out using the standard UVOT analysis tools available in the HEASoft v6.35 package. First, we co-added all the individual sky images from multiple observations using the tool \texttt{uvotimsum}, which creates a single, stacked deep image for better visualization of both the source and the background. We used a circular source aperture of radius 5$^{\prime\prime}$ centered on the source, and a nearby source-free circular region of radius 20$^{\prime\prime}$ to estimate the background. We then used the \texttt{uvotmaghist} tool to perform sequential aperture photometry on the co-added images and generate the light curve. This tool leverages the \texttt{uvotsource} tool to compute net source count rate, apply necessary calibrations and corrections, and convert the count rates into fluxes.

\section{X-ray Spectral Analysis}
\label{sec:x-ray_spec}

\begin{figure*}
    \begin{interactive}{animation}{xmm_pn_rgs_animation_video.mp4}
    \includegraphics[width=0.975\textwidth]{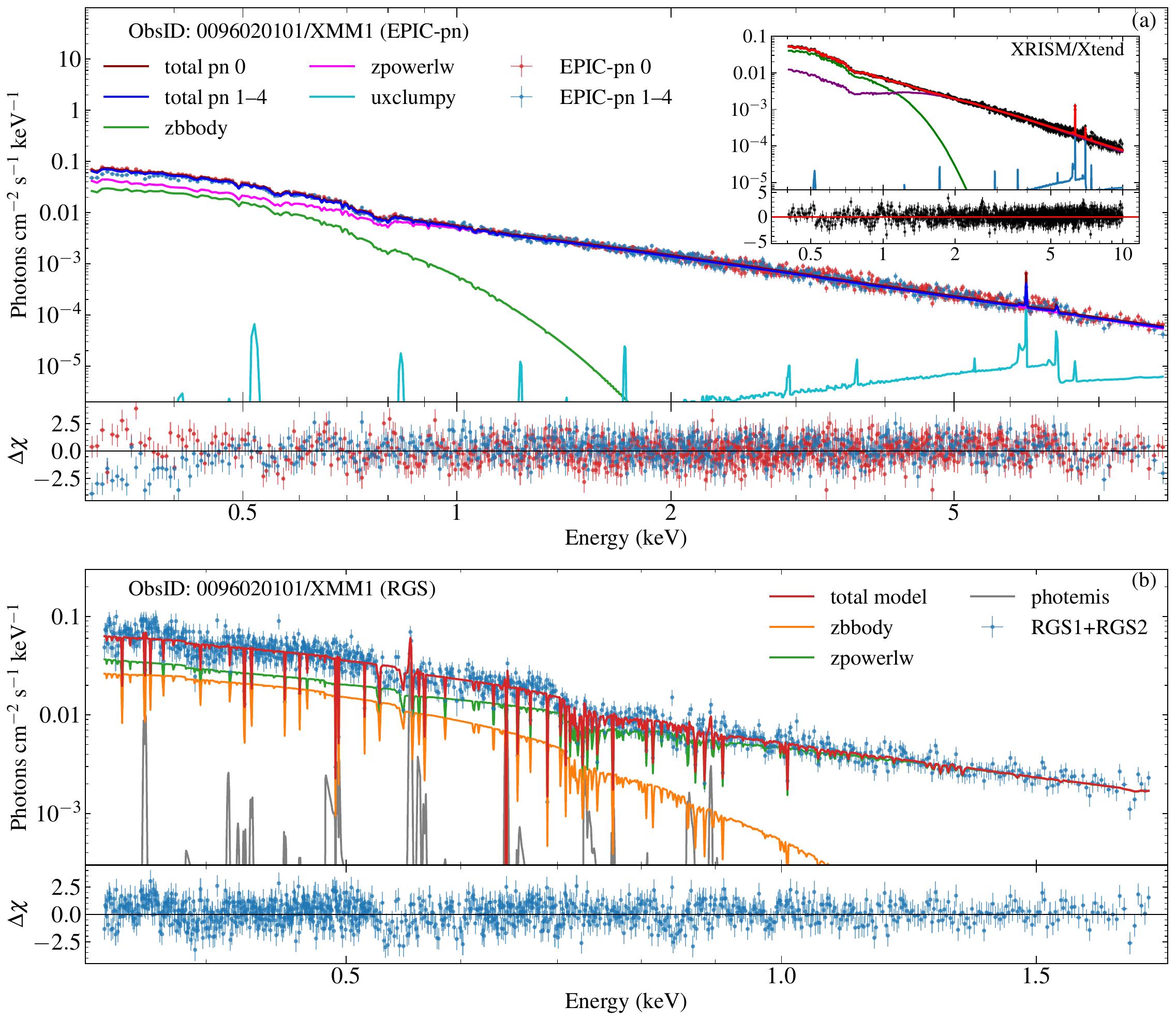}
    \end{interactive}
    \caption{Broadband X-ray spectra of Mrk 766 from the \xmm\ and \xrism\ observations. \textbf{(a)} EPIC-pn spectrum (0.3--10 keV) for the XMM1 epoch. Red and blue data points represent single-pixel (pattern 0) and double-pixel (patterns 1--4) events, respectively. The solid curve shows the best-fit continuum model consisting of a soft excess component (zbbody), a primary power-law continuum (zpowerlw), and torus reflection (uxclumpy), modified by Galactic absorption and a partially covering ionized absorber. Residuals with respect to the best-fit model are shown in the lower panel. The \textbf{inset} displays the XRISM/Xtend spectrum in 0.4--10 keV band fitted with the same phenomenological model. \textbf{(b)} Combined RGS1+RGS2 spectrum (0.34--1.8 keV) with the best-fit model including the continuum components, two warm absorber phases, and a photoionized emission component (photemis). The bottom panel shows the residuals relative to the best-fit model. The online animated version of this figure cycles through all nine XMM-Newton epochs (XMM1--XMM9).}
    \label{fig:EPIC-pn_rgs_anim}
\end{figure*}

\subsection{Continuum Analysis}
\label{subsec:epn}

    We modeled the EPIC-pn 0.3-10 keV X-ray spectra for all nine epochs of \textit{XMM-Newton} observations of Mrk~766 using the phenomenological model \texttt{constant $*$ TBabs $*$ zxipcf $*$ (zbbody + zpowerlw + uxclumpy)} implemented in \texttt{XSPEC v12.15.0} \citep{Arnaud_Xspec}. Galactic absorption was modeled using the Tuebingen–Boulder model (\texttt{TBabs}; \citealt{Wilms_2000}) with the hydrogen column density fixed at the Galactic value toward Mrk~766 ($N_{\mathrm{H, Gal}} = 1.85 \times 10^{20}~\mathrm{cm}^{-2}$; \citealt{HI4PI_Nh}). A partially covering ionized absorber (\texttt{zxipcf}) was employed to model variable absorption from warm ionized gas intrinsic to the source. The \texttt{zbbody} component provides a phenomenological description of the soft excess below $\sim$2 keV. Although the soft excess may arise from warm Comptonization or relativistic reflection \citep{Crummy_2006, Done_2012, Petrucci_2018}, we do not assign a direct physical interpretation to the blackbody temperature, which was allowed to vary to capture spectral variability. The \textsc{Uxclumpy}\footnote{\textsc{Uxclumpy}: \url{https://github.com/JohannesBuchner/xars/blob/master/doc/uxclumpy.rst}} \citep{Buchner_uxcl} additive model accounts for contributions from reprocessed X-ray emission in a clumpy torus. While the component is more prominent in heavily obscured sources, it was included to test for any torus-like reflection signatures or scattered light. This combined model effectively calculates the primary AGN continuum, soft excess, and complex absorption features. All nine spectra were fitted using this model, with parameters tied or untied as appropriate based on physical expectations and statistical tests. The model provides statistically acceptable fits across all observations. Figure~\ref{fig:EPIC-pn_rgs_anim} presents the best-fit model overlaid on the data for each epoch and the model best-fit parameters are listed in Table \ref{tab:summary_table}.

    To ensure consistency in cross-mission comparison, we also analyzed the 0.3–10 keV spectra of \xrism-Xtend using the same baseline phenomenological model \texttt{TBabs $*$ zxipcf $*$ (zbbody + zpowerlw + uxclumpy)} adopted for the \textit{XMM-Newton}/EPIC-pn data. The model parameters were allowed to vary freely, with no prior constraints imposed from the EPIC-pn fits as the \xrism\ observation is not contemporaneous with the \textit{XMM-Newton} observations. The primary continuum exhibits a photon index of $\Gamma = 2.28 \pm 0.02$ fully consistent with that derived from the \textit{XMM-Newton}/EPIC-pn spectra (see Table~\ref{tab:summary_table}). The inclusion of a partially covering ionized absorber significantly improves the fit, and also the ionization parameter value is consistent with the EPIC-pn fits, reaffirming the presence of warm absorption intrinsic to the source. The phenomenological model provides a statistically acceptable fit to the Xtend data shown in Figure~\ref{fig:EPIC-pn_rgs_anim} inset, and the best-fit parameter values listed in Table~\ref{tab:summary_table} are broadly consistent with those obtained from EPIC-pn, supporting a stable spectral shape.

\subsection[Fe K alpha Emission Line Profile]{\fekalpha\ Emission Line Profile}
\label{subsec:xrism_spectra}

\begin{figure*}[hbt!]
    \centering
    \includegraphics[width=\columnwidth]{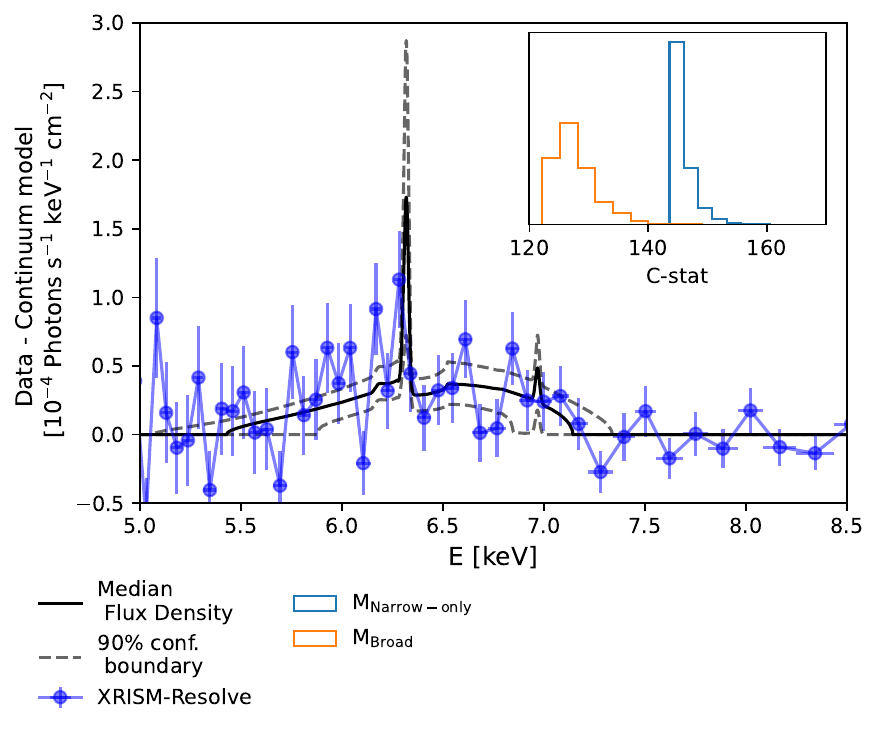}
    \includegraphics[width=\columnwidth]{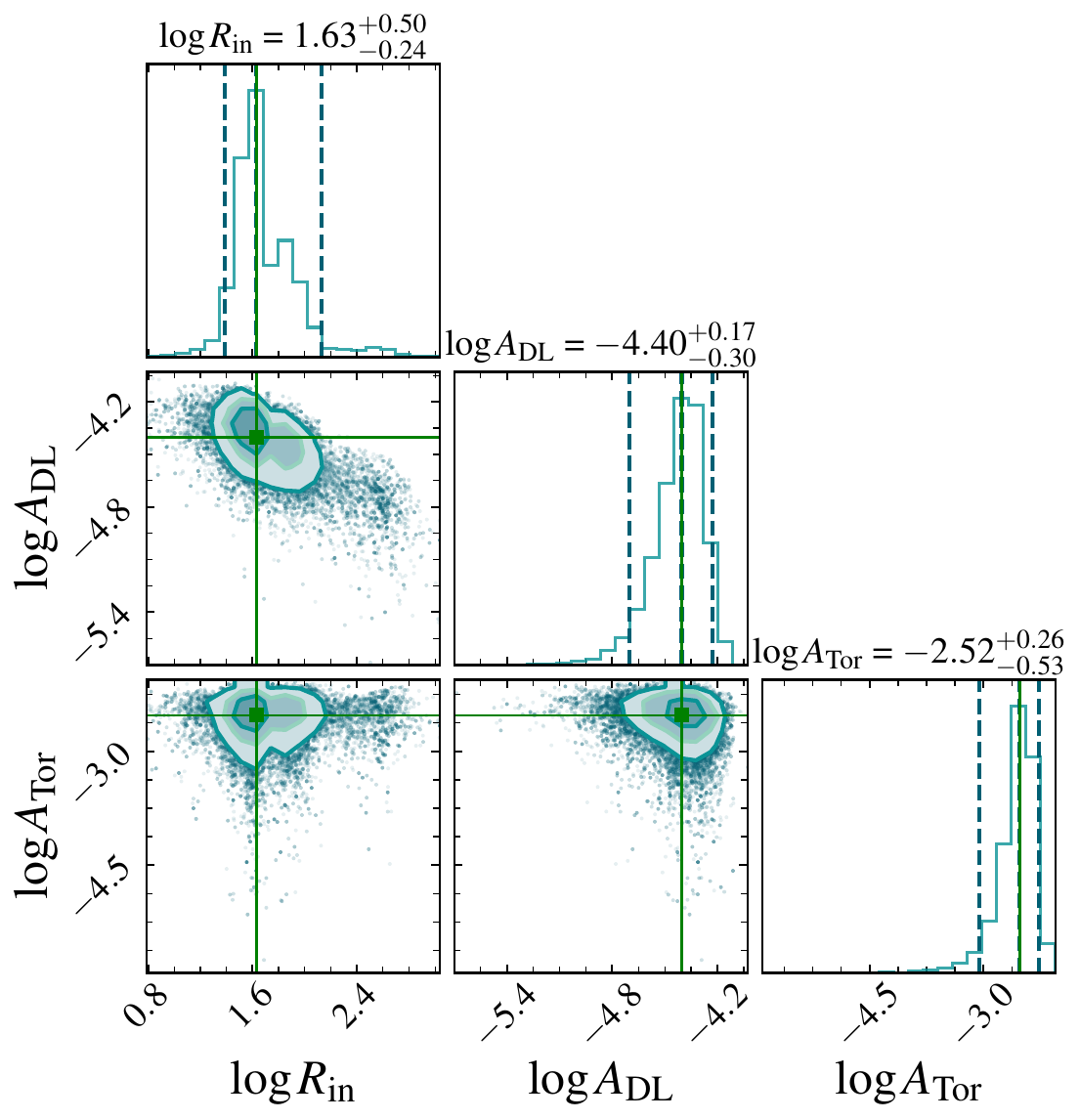}
    \caption{\textbf{(Left)} \xrism-Resolve spectrum in the 5--8.5 keV energy band fitted with the $M_{\rm broad}$ model, consisting of a narrow torus component (\texttt{MYTorus}) and a relativistic diskline representing emission from the inner accretion disk. The \textbf{inset} shows the distribution of the C-statistic values obtained from the MCMC exploration, indicating that the $M_{\rm broad}$ model provides a statistically improved fit compared to a narrow-line-only model. \textbf{(Right)} Posterior distributions and parameter correlations derived from the MCMC analysis for selected parameters: the inner radius of the \fekalpha\ emitting region ($R_{\rm{in}}$), the diskline normalization ($A_{\rm{DL}}$), and the torus normalization ($A_{\rm{Tor}}$).}
    \label{fig:xrism_resolve_xtend_Feline}
\end{figure*}

    The effective area of the \xrism-Resolve instrument in its GV closed configuration corresponds to a `usable' energy range of 3.0--10~keV.
    This enables only a deep study of the \fekalpha\ line region to date for this source.
    We fit the data with two different line models superposed on the same continuum, \texttt{M$_{\rm cont}$ = TBabs$*$(zbbody + powerlaw)}.
    Our final models are thus, \texttt{M$_{\rm Narrow-only}$ = M$_{\rm cont}$ + gsmooth(MYTORUS\_LINE)} and \texttt{M$_{\rm Broad}$ = M$_{\rm cont}$ + gsmooth(MYTORUS\_LINE) + diskline}.
    The \texttt{diskline} \citep{Fabian_1989} component accounts for relativistic Fe K$\alpha$ emission from the inner accretion disk and the \texttt{MYTorus} \citep{Murphy_Yaqoob_2009} is a model for X-ray reprocessing in a toroidal geometry. Our goal is to check whether the \fekalpha\ component exhibits other broader components other than the torus contribution.
    We also tested whether ionized Fe emission lines, specifically Fe\,\textsc{xxv} (6.70\,keV) and Fe\,\textsc{xxvi} (6.97\,keV), are statistically required by adding two \texttt{zgauss} components to the baseline model. The improvement in fit statistic is not statistically significant, and we therefore retain the baseline model.
    Here we avoid an explicit reflection component featuring a K-edge \citep[e.g., \texttt{relxill};][]{Garcia_2014}, since previous modeling (e.g. \citealt{Leighly_1996ApJ, Nandra_1997ApJ, Pounds_2003MNRAS, Miller_2007, Buisson_2018, Mochizuki_2023}) have shown that because of the weak reflection component normalization, this does not affect the line complex significantly. In this case, the spectrum can alternatively be explained by combinations of variable absorption and narrow/broad emission components without requiring a strong reflection hump. We also adopted both models, \texttt{M$_{\rm Narrow-only}$} and \texttt{M$_{\rm Broad}$}, for the analyses of the \fekalpha\ complex in \xrism-Xtend 0.3–10 keV spectra.

\subsection{Outflow based on RGS}
\label{subsec:rgs}

    We modeled the data from the Reflection Grating Spectrometer (RGS; \citealt{Herder_2001}) X-ray spectra in the 0.34--1.8~keV band for all nine epochs of \textit{XMM-Newton} observations of Mrk~766 using the phenomenological model \texttt{TBabs$*$warmabs(1)$*$warmabs(2)$*$(zbbody + zpowerlw + photemis)} which accounts for both the intrinsic source emission and the effects of ionized absorption. We employed two distinct warm absorbers 
    (\texttt{warmabs}\footnote{\texttt{Warmabs} model: \url{https://heasarc.gsfc.nasa.gov/docs/software/xstar/docs/sphinx/xstardoc/docs/build/html/index.html}} model; \citealt{Kallman_2001, Kallman_2004}) 
    to model absorption from multiple ionization phases, allowing the column density ($N_{\mathrm{H}}$), ionization parameter ($\log \xi$), and turbulent velocities to vary independently. Each \texttt{warmabs} component represents a distinct ionized absorber along the line of sight and the photons pass through both \texttt{warmabs} components in sequence each one affecting the spectra differently depending upon the ionization of the WAs. We also included \texttt{photemis} model which is an additive emission component, used to model discrete narrow emission lines, likely originating from photoionized plasma in the narrow-line region or outflowing material. Figure~\ref{fig:EPIC-pn_rgs_anim} presents the best-fit model overlaid on the data for each epoch and the model best-fit parameters are listed in Table \ref{tab:summary_table}.

\section{Results}
\label{sec:results}


\begin{splitdeluxetable*}{cccccccccBcccccccccc}
\tablecaption{Summary of EPIC-pn, RGS, \xrism-Xtend and Resolve spectral fitting: 2--10 keV X-ray continuum flux, photon index ($\Gamma$), soft-excess temperature, partial covering absorber and warm absorber properties, and \fekalpha\ line properties. \label{tab:summary_table}}
\tablewidth{0pt}
\tablehead{
\colhead{Obs Date} & \colhead{Obs ID} & \colhead{Abbrv.} & 
\multicolumn{2}{c}{X-ray Continuum} & 
\colhead{kT\tablenotemark{b}} & 
\multicolumn{3}{c}{\texttt{zxipcf}} &
\multicolumn{4}{c}{\texttt{warmabs}\tablenotemark{d}} &
\colhead{\texttt{photemis}} &
\multicolumn{3}{c}{$\rm{M_{Broad}}$} &
\multicolumn{2}{c}{$\rm Fe~K\alpha$ complex flux\tablenotemark{e}} \\
\cline{4-5} \cline{7-9} \cline{10-13} \cline{15-17} \cline{18-19}
\colhead{(yyyy-mm-dd)} & & & 
\colhead{$F_{\rm 2-10~keV}$\tablenotemark{a}} & 
\colhead{$\Gamma$} & 
\colhead{(keV)} & 
\colhead{$N_{\rm H}$\tablenotemark{c}} &
\colhead{Cov$_{\rm frac}$} &
\colhead{$\log \xi$} &
\colhead{$N_{\rm H}^{(1)}$} &
\colhead{$\log \xi^{(1)}$} &
\colhead{$N_{\rm H}^{(2)}$} &
\colhead{$\log \xi^{(2)}$} &
\colhead{$\log \xi^{(3)}$} &
\colhead{$\log R_{\rm{in}}$} & 
\colhead{$\log A_{\rm{DL}}$} & 
\colhead{$\log A_{\rm{Tor}}$} & 
\colhead{$\rm{F_{Broad}}$} & 
\colhead{$\rm{F_{Narrow}}$}
}
\colnumbers
\startdata
2000-05-20 & 0096020101 & XMM1 & $1.527 \pm 0.020$ & $2.130 \pm 0.023$ & $0.107 \pm 0.002$ & $0.516 \pm 0.061$ & $0.580 \pm 0.021$ & $0.905 \pm 0.069$ & $-0.65^{+0.03}_{-0.03}$ & $0.96^{+0.06}_{-0.09}$ & $-2.07^{+0.18}_{-0.35}$ & $-1.02^{+0.38}_{-0.32}$ & $1.23^{+0.28}_{-0.47}$ & $1.72^{+0.14}_{-0.15}$ & $-4.63^{+0.10}_{-0.14}$ & $-2.19^{+0.14}_{-0.20}$ & $2.367^{+1.030}_{-1.044}$ & $1.356^{+0.832}_{-0.763}$ \\
2001-05-20 & 0109141301 & XMM2 & $2.410 \pm 0.015$ & $2.260 \pm 0.013$ & $0.126 \pm 0.002$ & $0.524 \pm 0.033$ & $0.646 \pm 0.009$ & $0.764 \pm 0.028$ & $-0.85^{+0.07}_{-0.07}$ & \textcolor{magenta}{\nodata} & $-0.67^{+0.01}_{-0.02}$ & $0.65^{+0.04}_{-0.04}$ & $1.99^{+0.10}_{-0.05}$ & $1.76^{+0.10}_{-0.09}$ & $-4.52^{+0.06}_{-0.06}$ & $-2.11^{+0.11}_{-0.15}$ & $3.150^{+0.723}_{-0.761}$ & $1.653^{+0.817}_{-0.829}$ \\
2005-05-23 & 0304030101 & XMM3 & $0.712 \pm 0.017$ & $1.920 \pm 0.056$ & $0.096 \pm 0.002$ & $7.460 \pm 0.690$ & $0.596 \pm 0.046$ & $1.607 \pm 0.163$ & $-0.94^{+0.06}_{-0.06}$ & $-1.28^{+0.22}_{-0.20}$ & $0.29^{+0.02}_{-0.02}$ & \textcolor{magenta}{\nodata} & $1.67^{+0.10}_{-0.13}$ & $2.21^{+0.15}_{-0.14}$ & $-4.79^{+0.06}_{-0.06}$ & $-3.11^{+0.28}_{-0.46}$ & $1.641^{+0.039}_{-0.038}$ & $0.156^{+0.135}_{-0.138}$ \\
2005-05-25 & 0304030301 & XMM4 & $1.113 \pm 0.029$ & $2.082 \pm 0.017$ & $0.104 \pm 0.001$ & $0.622 \pm 0.051$ & $0.575 \pm 0.016$ & $0.858 \pm 0.056$ & $-0.50^{+0.12}_{-0.13}$ & \textcolor{magenta}{\nodata} & $-0.86^{+0.07}_{-0.08}$ & $0.35^{+0.18}_{-0.26}$ & $1.78^{+0.13}_{-0.16}$ & $2.96^{+0.21}_{-0.24}$ & $-5.04^{+0.09}_{-0.12}$ & $-2.29^{+0.12}_{-0.16}$ & $0.944^{+0.388}_{-0.404}$ & $1.087^{+0.573}_{-0.558}$ \\
2005-05-27 & 0304030401 & XMM5 & $1.375 \pm 0.011$ & $2.124 \pm 0.015$ & $0.113 \pm 0.002$ & $0.520 \pm 0.034$ & $0.646 \pm 0.012$ & $0.813 \pm 0.039$ & $-0.73^{+0.03}_{-0.03}$ & $0.53^{+0.08}_{-0.08}$ & $-0.93^{+0.23}_{-0.16}$ & \textcolor{magenta}{\nodata} & $1.34^{+0.26}_{-0.20}$ & $1.73^{+0.07}_{-0.08}$ & $-4.64^{+0.06}_{-0.07}$ & $-2.71^{+0.22}_{-0.37}$ & $2.335^{+0.557}_{-0.564}$ & $0.432^{+0.425}_{-0.362}$ \\
2005-05-29 & 0304030501 & XMM6 & $1.740 \pm 0.011$ & $2.125 \pm 0.010$ & $0.114 \pm 0.002$ & $0.610 \pm 0.045$ & $0.696 \pm 0.023$ & $1.278 \pm 0.033$ & $-0.91^{+0.04}_{-0.04}$ & $0.68^{+0.09}_{-0.07}$ & $-0.54^{+0.06}_{-0.07}$ & \nodata & $1.64^{+0.25}_{-0.22}$ & $1.70^{+0.10}_{-0.08}$ & $-4.70^{+0.07}_{-0.09}$ & $-2.40^{+0.13}_{-0.18}$ & $2.034^{+0.602}_{-0.622}$ & $0.845^{+0.443}_{-0.497}$ \\
2005-05-31 & 0304030601 & XMM7 & $1.505 \pm 0.041$ & $2.115 \pm 0.017$ & $0.108 \pm 0.002$ & $0.696 \pm 0.023$ & $0.592 \pm 0.015$ & $0.832 \pm 0.048$ & $-0.79^{+0.03}_{-0.04}$ & $0.63^{+0.07}_{-0.11}$ & $-1.09^{+0.25}_{-0.23}$ & $1.89^{+0.09}_{-0.10}$ & $1.49^{+0.28}_{-0.19}$ & $1.94^{+0.13}_{-0.10}$ & $-4.71^{+0.07}_{-0.09}$ & $-2.72^{+0.25}_{-0.44}$ & $1.971^{+0.653}_{-0.628}$ & $0.397^{+0.316}_{-0.340}$ \\
2005-06-03 & 0304030701 & XMM8 & $1.180 \pm 0.055$ & $2.072 \pm 0.029$ & $0.102 \pm 0.003$ & $0.561 \pm 0.113$ & $0.595 \pm 0.059$ & $1.138 \pm 0.125$ & $-0.93^{+0.11}_{-0.10}$ & $0.58^{+0.11}_{-0.10}$ & $-0.76^{+0.08}_{-0.07}$ & $1.88^{+0.11}_{-0.09}$ & $1.43^{+0.23}_{-0.22}$ & $2.22^{+0.91}_{-0.60}$ & $-4.99^{+0.28}_{-0.33}$ & $-2.41^{+0.20}_{-0.33}$ & $1.033^{+1.534}_{-0.806}$ & $0.803^{+0.742}_{-0.639}$ \\
2015-07-05 & 0763790401 & XMM9 & $1.442 \pm 0.052$ & $2.113 \pm 0.026$ & $0.112 \pm 0.004$ & $0.595 \pm 0.059$ & $0.569 \pm 0.025$ & $0.669 \pm 0.072$ & $-0.95^{+0.11}_{-0.14}$ & $0.98^{+0.15}_{-0.19}$ & $-0.91^{+0.20}_{-0.20}$ & $1.91^{+0.10}_{-0.05}$ & $3.30^{+0.12}_{-0.11}$ & $2.13^{+0.31}_{-0.37}$ & $-4.89^{+0.15}_{-0.21}$ & $-2.57^{+0.26}_{-0.47}$ & $1.347^{+0.897}_{-0.900}$ & $0.576^{+0.828}_{-0.494}$ \\
\tableline
2024-06-24 & 300007010 & Xtend & $2.127^{+0.011}_{-0.022}$ & $2.277 \pm 0.019$ & $0.132 \pm 0.001$ & $1.081 \pm 0.046$ & $0.896 \pm 0.007$ & $0.339 \pm 0.007$ & \nodata & \nodata & \nodata & \nodata & \nodata & $1.65^{+0.21}_{-0.20}$ & $-4.60^{+0.15}_{-0.22}$ & $-1.94^{+0.15}_{-0.24}$ & $2.609^{+1.105}_{-1.040}$ & $2.420^{+1.015}_{-1.051}$ \\
2024-06-24 & 300007010 & Resolve & \nodata & \nodata & \nodata & \nodata & \nodata & \nodata & \nodata & \nodata & \nodata & \nodata & \nodata & $1.63^{+0.50}_{-0.24}$ & $-4.40^{+0.17}_{-0.30}$ & $-2.52^{+0.26}_{-0.53}$ & $2.626 \pm 1.076$ & $2.407 \pm 0.996$ \\
\enddata
\tablenotetext{a}{Observed 2--10 keV flux, in units of $10^{-11}$ erg cm$^{-2}$ s$^{-1}$.}
\tablenotetext{b}{Soft-excess temperature (\texttt{zbbody} model).}
\tablenotetext{c}{Partial covering absorber model. Column density in units of $10^{22}$ cm$^{-2}$.}
\tablenotetext{d}{Warm absorber model. $N_{\rm H}^{(1)}$ and $N_{\rm H}^{(2)}$ are absorption column densities in log scale ($10^{22}$ cm$^{-2}$). For epochs where the ionization parameter ($\log \xi^{(1)}$, $\log \xi^{(2)}$) posterior is unconstrained (flat distribution); value not reported.}
\tablenotetext{e}{\fekalpha\ complex flux in units of $10^{-13}$ erg cm$^{-2}$ s$^{-1}$.}
\end{splitdeluxetable*}


\subsection[Fe K alpha Line Evolution]{\fekalpha\ Line Evolution} 
\label{subsec:Fe_K_line}

\begin{figure}
    \centering
    \includegraphics[width=\columnwidth]{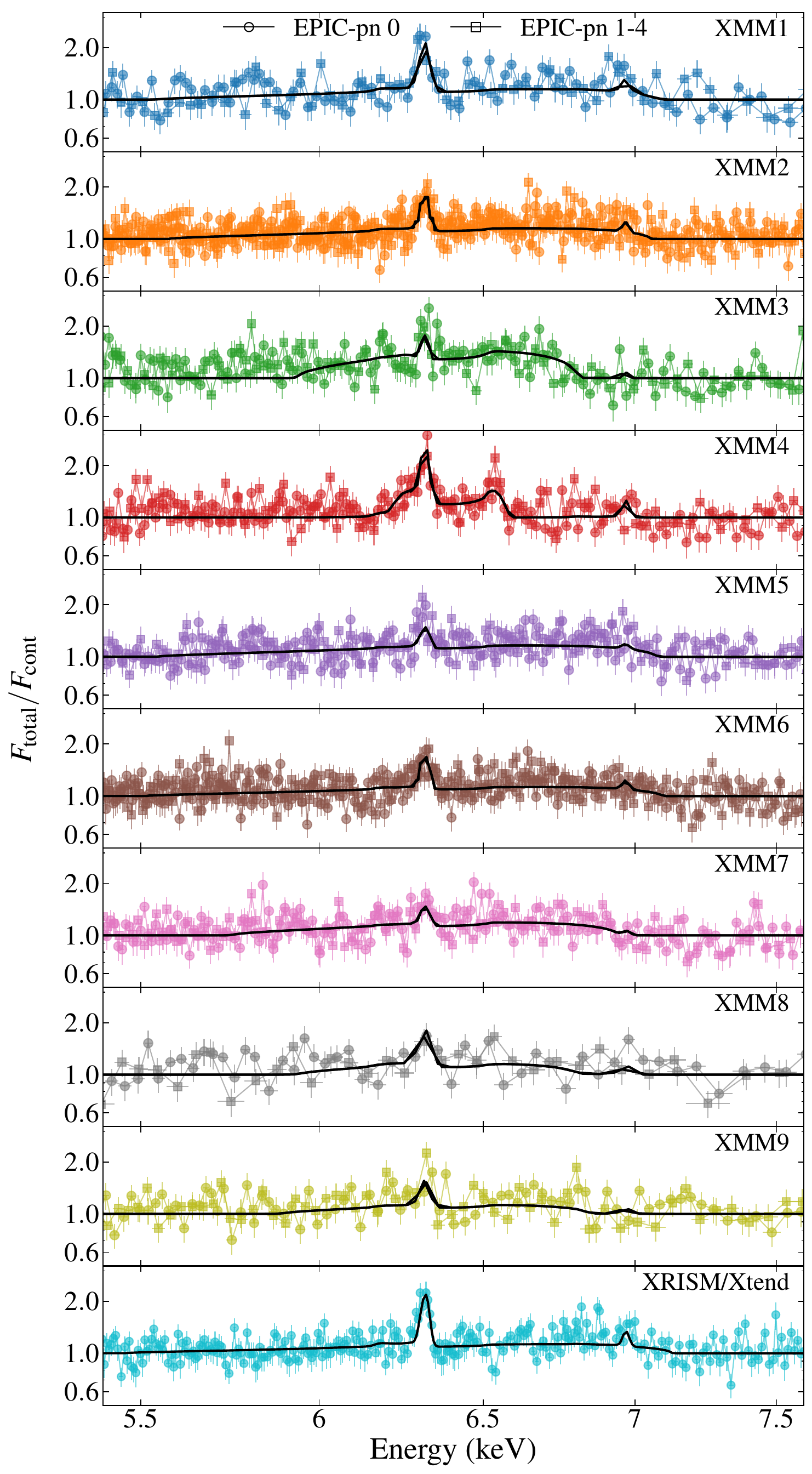}
    \caption{Ratio spectra showing the evolution of the \fekalpha\ line complex relative to the continuum across nine epochs of \xmm\ observations and the \xrism-Xtend observation. Each panel shows the EPIC-pn data (single- and double-pixel events) normalized with the best-fit continuum model, highlighting the \fekalpha\ emission feature around 6.4 keV. The black curve represents the best-fit model including both narrow torus emission and a relativistic diskline component.}
    \label{fig:Feline_evolution}
\end{figure}

\begin{figure}
    \centering
    \includegraphics[width=\columnwidth]{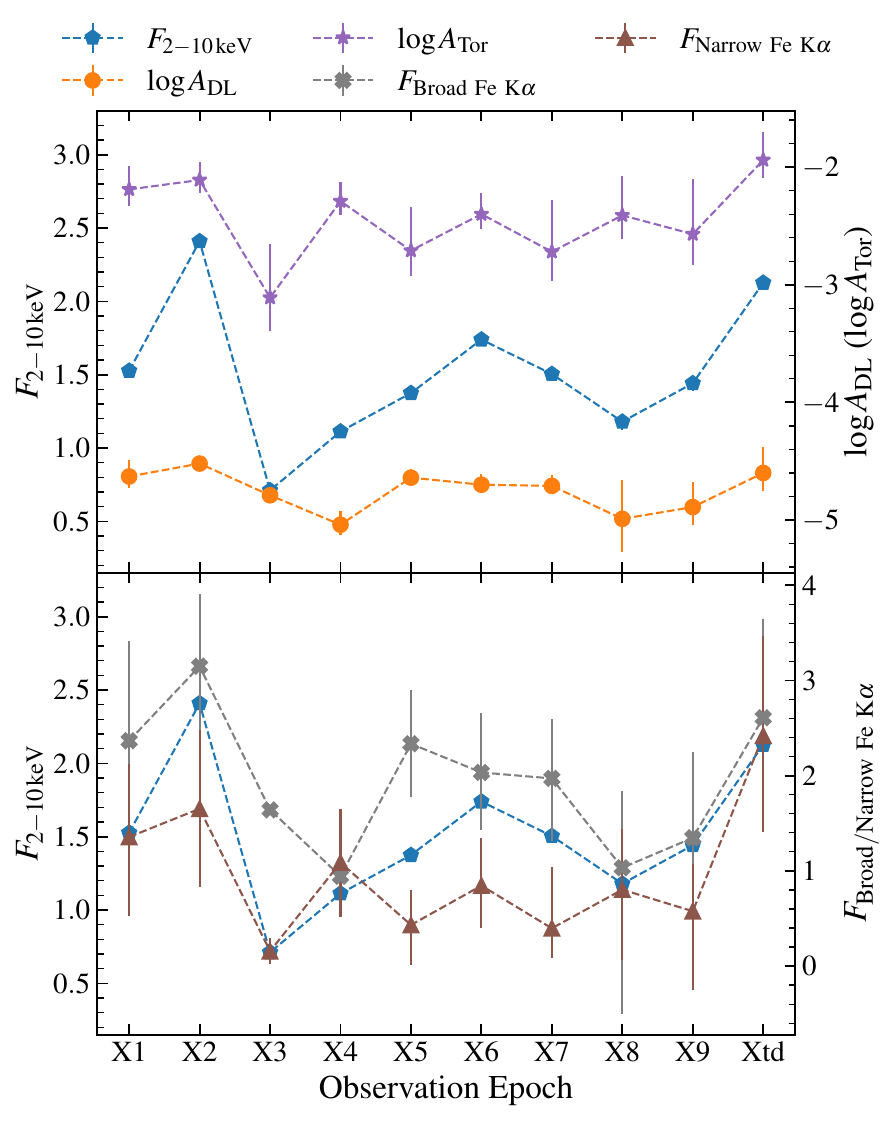}
    \caption{\textbf{Top panel:} Variation of the 2--10 keV continuum flux together with the logarithmic normalizations of the diskline component ($\log A_{\rm DL}$) and the torus fluorescence component ($\log A_{\rm Tor}$) across different epochs of \xmm\ and \xrism-Xtend observation. \textbf{Bottom panel:} Variation of the broad (F$_{\rm Broad}$) and narrow (F$_{\rm Narrow}$) \fekalpha\ line fluxes.
    The \texttt{diskline} normalization ($A_{\rm DL}$) and broad \fekalpha\ line flux closely follow the continuum flux.}
    \label{fig:Feline_params_epochs}
\end{figure}
    
    The line profile as captured by a \xrism\ observation from 2024 displays a narrow 6.4 keV core together with excess emission on both sides of the line, which cannot be captured by the \texttt{M$_{\rm Narrow-only}$} model alone. MCMC exploration of the parameter space confirms that the \texttt{M$_{\rm Broad}$} model provides a statistically better description of the data over the \texttt{M$_{\rm Narrow-only}$} model (Section~\ref{subsec:Fe_K_line}, Figure~\ref{fig:xrism_resolve_xtend_Feline}), as indicated by distribution of the fit-statistic of the samples, where it is distinctly shifted toward lower C-stat values for the \texttt{M$_{\rm Broad}$} model, in which the broad \texttt{diskline} is include. 
    Thus, we adopted the \texttt{M$_{\rm Broad}$} model for all other datasets of \xrism-Xtend data and nine epochs of \xmm\ \epn\ data to track the evolution of the Fe K$\alpha$ complex across multiple epochs of \textit{XMM-Newton} observations. The high spectral resolution of Resolve allows us to recover tight constraints on the broadened component, the inner radius is $\log R_{\rm{in}} = 1.63^{+0.50}_{-0.24}~r_{\mathrm{g}}$, and the normalizations of the disk and torus components are $\log A_{\rm{DL}} = -4.40^{+0.17}_{-0.30}$ and $\log A_{\rm{Tor}} = -2.52^{+0.26}_{-0.53}$, respectively. These results indicate that the broadened \fekalpha\ emission arises from radii of tens of gravitational radii, which is large enough to avoid extreme gravitational redshift, but small enough to require relativistic broadening, while the narrow core originates in a more distant structure.
    
    The \xrism-Xtend spectrum, though lower in spectral resolution, provides an important consistency check. Fitting the Xtend data with the same modeling produces identical constraints on the line structure, with $\log R_{\rm{in}} = 1.65^{+0.20}_{-0.21}~r_{\mathrm{g}}$ and similar disk and torus normalizations. The agreement between the Resolve and Xtend posteriors indicates that the recovered parameters are robust to instrumental differences and do not depend on subtle degeneracy between the baseline continuum and emission line normalizations.
    
    The \texttt{diskline} component represents a possible broad component in the \fekalpha\ line complex. However, we caution the reader that the model should not be interpreted literally. We are aware that a physically motivated model like \texttt{relxill} is more appropriate for modeling a broad line coupled with relativistic reflection; however, our data quality and photon statistics around the 6.4~keV \fekalpha\ complex are not adequate for constraining the other free parameters (e.g. black hole spin, ionization parameter, coronal height) that are included in the model. The \texttt{diskline} model with `optimum' number of parameters will capture the extent of the evolving broad line-emitting region and effectively reflect the trend with respect to the continuum, demonstrating that the line-emitting region evolves.

    The nine \epn\ epochs reveal a coherent long-term picture of \fekalpha\ evolution. A 6.4 keV narrow line core persists across all epochs, consistent with a stable reflector such as the torus or outer BLR. The prominence of any broader component, however, varies with time. In several epochs the \epn\ spectra show weak but discernible wings around the 6.4 keV core shown in Figure~\ref{fig:Feline_evolution}, favouring the \texttt{M$_{\rm Broad}$} model, whereas in others the narrower \texttt{MYTorus} component provides an adequate description. This variability likely reflects changes in the illuminating continuum, partial-covering absorption, or coronal geometry, which can modulate the relative visibility of inner-disk reflection in CCD data. The calorimeter data from Resolve demonstrate that a broadened component is indeed present, even when CCD spectra alone cannot unambiguously isolate it.

\begin{figure*}
    \centering
    \includegraphics[width=\textwidth]{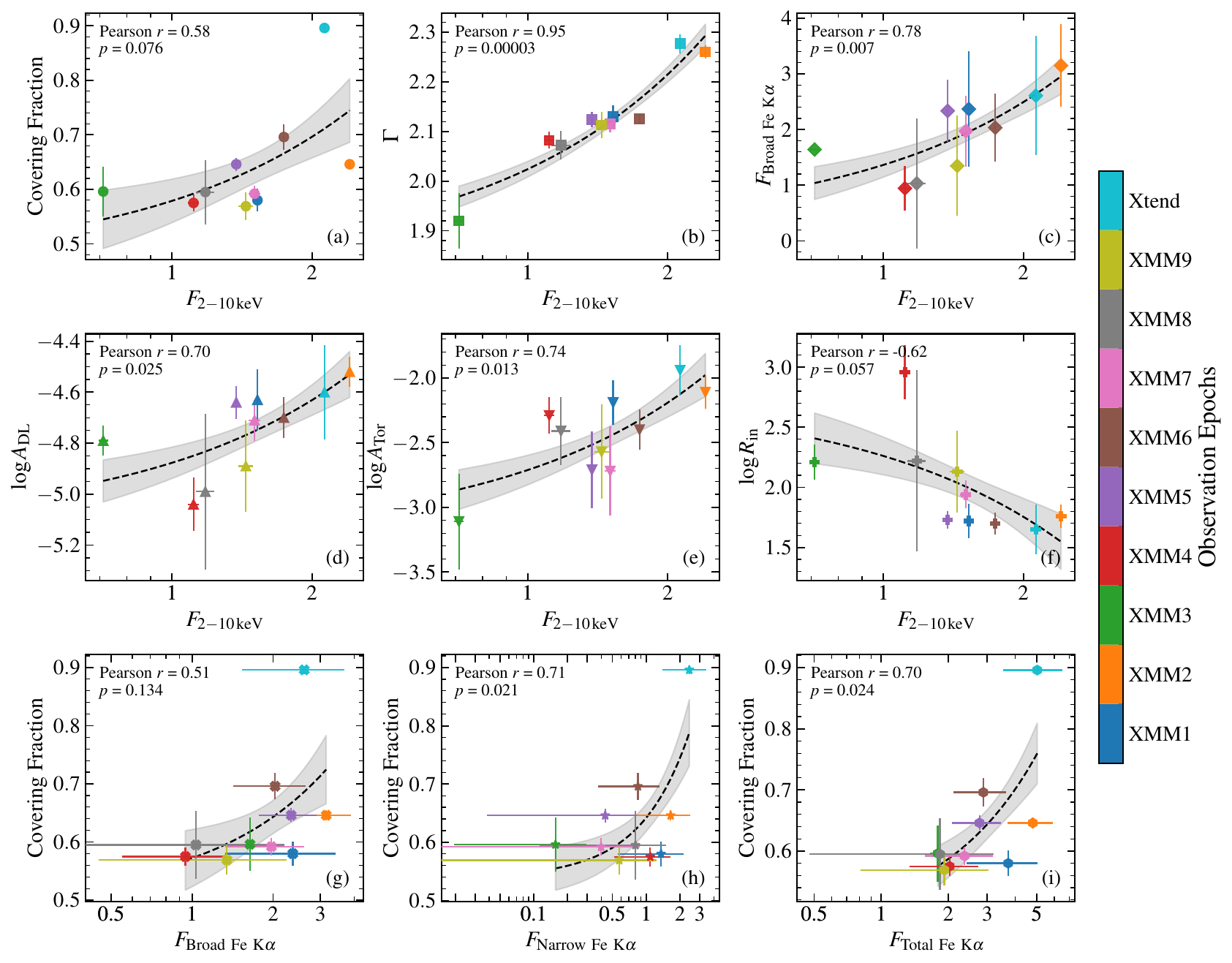}
    \caption{First and second rows of panels show the 
    correlation between the intrinsic 2–10 keV continuum flux with the following parameters: 
    \textbf{(a)} covering fraction of the absorber (section~\ref{subsec:epn}), 
    \textbf{(b)} coronal power-law photon index ($\Gamma$),
    \textbf{(c)} the broad \fekalpha\ flux,
    \textbf{(d)} normalization of the \texttt{diskline} broad component ($A_{\rm DL}$),
    \textbf{(e)} normalization of the narrow \fekalpha\ ($A_{\rm Tor}$) component modeled with \texttt{MYTORUS} (section~\ref{subsec:xrism_spectra}), and
    \textbf{(f)} the inner radius ($R_{\rm in}$) of the broad \fekalpha\ emitter modeled with \texttt{diskline}, 
    and the third row of panel \textbf{(g, h, i)} shows the correlation between the \fekalpha\ broad, narrow, and total flux with the absorber covering fraction, respectively.
    In each subplot, the black dashed line represents the best-fit linear relation of the form $y=mx+c$ obtained using the least-squares regression, and the shaded colored region indicates the $1\sigma$ confidence interval on the best-fit linear regression line. The Pearson correlation coefficient ($r$) and the corresponding $p$-value are shown in each panel.
    The data covers all nine \xmm\ and the \xrism-Xtend observation.
    The continuum 2--10 keV flux values are in units of $10^{-11}$~erg~cm$^{-2}$~s$^{-1}$, and the \fekalpha\ line flux values are in units of $10^{-13}$~erg~cm$^{-2}$~s$^{-1}$. The regression is performed in linear space, while the horizontal axis of the subplots is shown in logarithmic scale for visualization.}
    \label{fig:correlation_plot}
\end{figure*}

    The temporal behaviour of the fitted model parameters across the nine EPIC-pn epochs is summarized in Figure \ref{fig:Feline_params_epochs}. The top panel in Figure \ref{fig:Feline_params_epochs} shows the evolution of the continuum flux in the 2--10 keV band together with the logarithmic normalizations of the diskline ($\log A_{\rm{DL}}$) and the torus fluorescence component ($\log A_{\rm{Tor}}$). A clear pattern emerges that $\log A_{\rm{DL}}$ tracks the continuum flux closely, indicating that the inner-disk \fekalpha\ emission responds promptly to changes in the illuminating continuum. This behaviour is consistent with a compact reprocessing region located at a distance of tens of gravitational radii, consistent with the MCMC constraints obtained from the \xrism\ spectra. By contrast, the \texttt{MYTorus} normalization ($\log A_{\rm{Tor}}$) is comparatively stable but also tracks the continuum in higher flux states, showing only modest variability across epochs, which supports its association with a distant reprocessor such as the torus or outer BLR, which is expected to respond on much longer timescales. The bottom panel of Figure \ref{fig:Feline_params_epochs} further shows the relationship between the continuum level and the line fluxes. The broad \fekalpha\ line component flux, $F_{\rm{Broad}}$, is strongest during epochs with enhanced continuum flux, and its amplitude generally decreases as the source dims, suggesting a positive response from the inner-disk emission. In contrast, the narrow flux, $F_{\rm{Narrow}}$, remains comparatively constant with only mild fluctuations, though $F_{\rm{Narrow}}$ is enhanced during stronger continuum fluxes. Figure~\ref{fig:correlation_plot} quantifies the relationship between the \fekalpha\ emission components and the intrinsic 2--10~keV continuum flux across the nine \xmm\ epochs. The broad \fekalpha\ component shows a clear positive correlation with the continuum flux (Pearson $r \simeq 0.78$, $p \simeq 0.007$), indicating that the strength of the broadened line increases during brighter states. In contrast, the narrow \fekalpha\ component exhibits weaker correlation with the continuum level. This dichotomy demonstrates that the two \fekalpha\ components respond on different timescales, with the broad component closely coupled to the instantaneous coronal emission, while the narrow component remains comparatively stable.

\subsection[Connection Between Partial Covering Variability and Fe K alpha Emission]{Connection Between Partial Covering Variability and \fekalpha\ Emission}
\label{subsec:partial_covr_variability}

    The covering fraction of the ionized absorber (\texttt{zxipcf}) displays a mild but systematic increase with the intrinsic 2--10 keV flux (Figure~\ref{fig:correlation_plot}, panel\textbf{(a)}), with Pearson $r = 0.58\, (p = 0.076)$. Although not statistically strong, this behaviour is consistent with luminosity-dependent changes in the geometry of the ionized outflow. In high-Eddington ratio AGN such as NLSy1s, radiation-hydrodynamic simulations predict that increasing accretion power strengthens radiation-driven disk winds and produces more vertically collimated outflow funnels \citep{Proga_2004, Proga_2007, Higgin_2014}. A more confined, vertically extended wind geometry increases the likelihood that clumpy ionized material intersects the line of sight, naturally leading to a higher effective covering fraction during brighter epochs. Observationally, luminous NLSy1s and quasars exhibit exactly this behaviour that enhanced accretion luminosity is associated with stronger or more coherent ionized winds \citep{Tombesi_2013, Parker_2017, Matzeu_2017, Parker_2018}.
    The positive correlation between covering fraction and continuum flux therefore provides a natural explanation for the multi-epoch \fekalpha\ line behaviour. During higher accretion states, radiation-driven disk wind models predict a more structured and vertically extended outflow. The observed increase in covering fraction with flux suggests a possible link between accretion state and absorber geometry. In such a scenario, changes in the disk illumination pattern may enhance the relativistic \fekalpha\ emission, consistent with the observed correlation between the broad component and the continuum. In lower-flux epochs, when the wind is less structured or the clumpy absorber geometrically covers less of the line of sight, the broad component weakens.

\subsection[Flux-dependent Fe K alpha Emission-radius and Disk–corona Geometry]{Flux-dependent \fekalpha\ Emission-radius and Disk–corona Geometry}
\label{Disk–corona_Geometry}

    Figure~\ref{fig:correlation_plot}, panel\textbf{(f)} shows a clear anti-correlation between the 2--10 keV continuum flux and the inferred \fekalpha\ line–emitting radius with a Pearson correlation coefficient of $r \sim -0.62$. Although the statistical significance is marginal ($p \sim 0.057$), the trend suggests that the characteristic reflection radius decreases systematically as the continuum brightens. This behavior is naturally expected if the illumination pattern of the accretion disk changes with the coronal emission state. In higher-flux states, the enhanced strength of the broad \fekalpha\ component suggests a more centrally concentrated illumination pattern, possibly due to changes in the coronal geometry (e.g., a more compact or lower-height corona), leading to stronger relativistic broadening and a smaller effective \fekalpha\ emission radius. In contrast, during lower-flux states, disk illumination appears less centrally concentrated, and the dominant contribution to the \fekalpha\ emission shifts to larger radii. Such flux-dependent changes in the reflection geometry are consistent with scenarios in which the corona varies in height, compactness, or radial extent above the disk, modifying the illumination profile across the inner accretion flow.

\subsection{UV--X-ray correlation}
\label{subsec:uv_x_corr}

\begin{figure*}[htb!]
    \centering
    \includegraphics[width=\columnwidth]{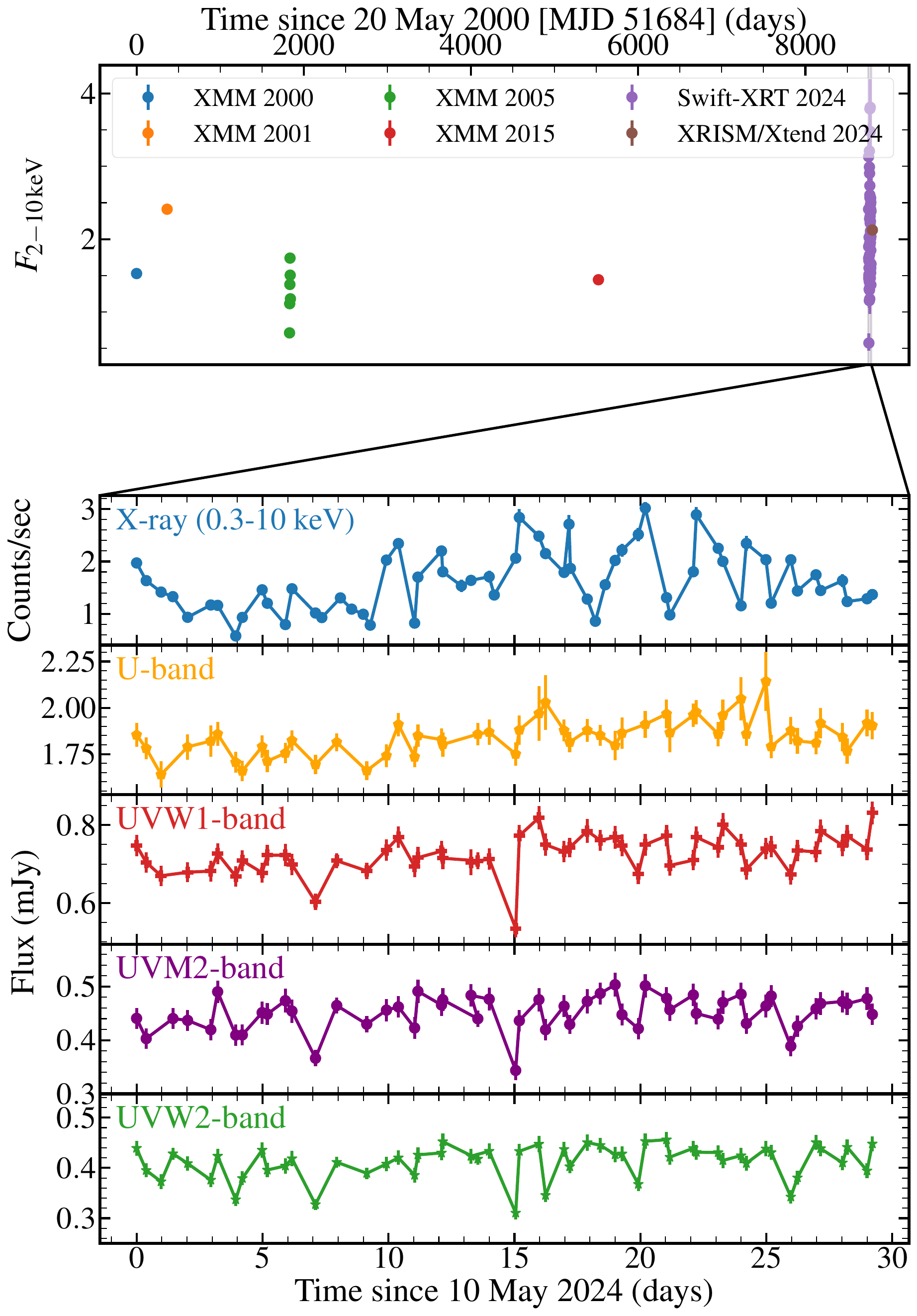}
    \includegraphics[width=\columnwidth]{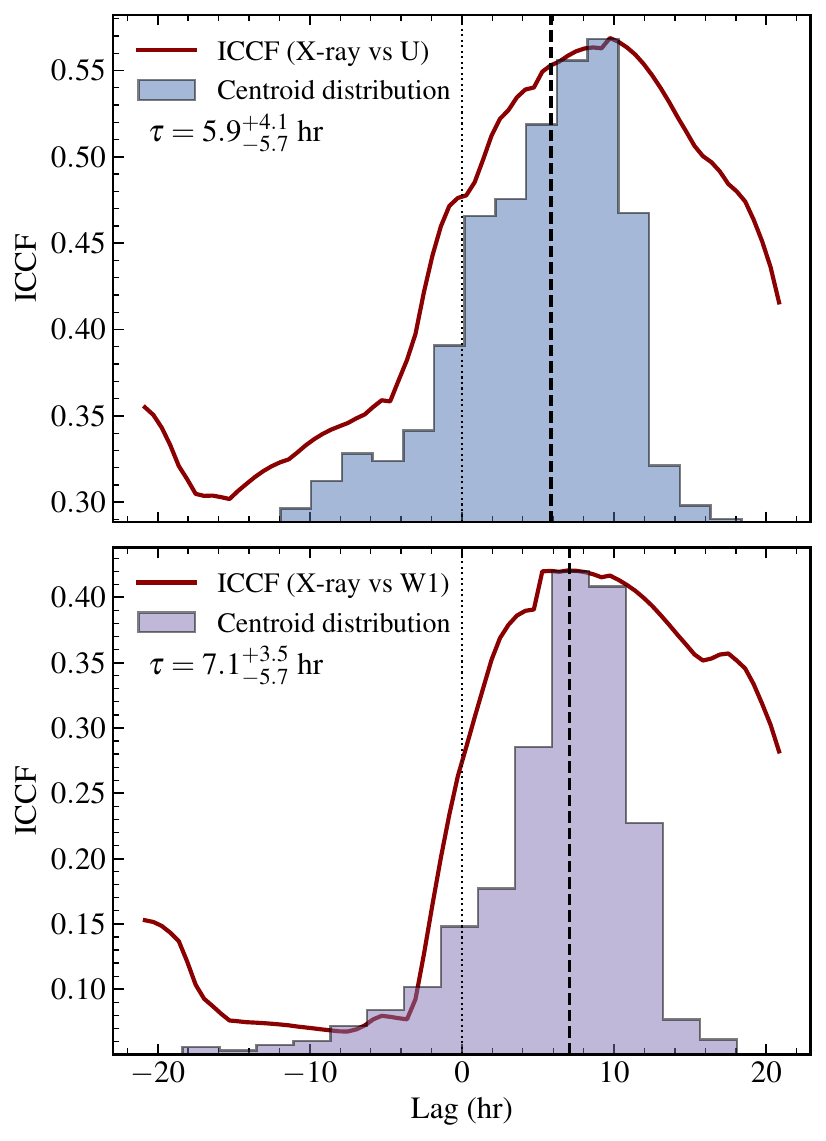}
    \caption{\textbf{Left panel:} Long-term and short-term variability of Mrk 766 in the X-ray and ultraviolet bands. X-ray and UV/optical light curves of Mrk~766. The top panel shows the 2--10 keV X-ray flux of the all \xmm\ epochs along with \swift-XRT and \xrism-Xtend observations, while the lower panels present the \textit{Swift}-XRT (0.3--10 keV) and UVOT (U, UVW1, UVW2, UVM2 bands) light curves of Mrk~766 obtained during May--June 2024. \textbf{Right panel:} Interpolated cross-correlation function (ICCF) between the \textit{Swift} X-ray and U-band and UVW1-band light curves shown using the red curve. The distribution of centroid lags from 5000 flux randomization/random subset selection (FR/RSS) realizations is shown using the histogram. Estimated lag from the FR/RSS centroid distribution indicates that the U-band variations lag by $5.9^{+4.1}_{-5.7}$ hours and UVW1-band by $7.1^{+3.5}_{-5.7}$ hours behind the X-ray variations. This is consistent with a reprocessing origin of the U/UVW1-band variations.}
    \label{fig:swift_lc_ICCF}
\end{figure*}

    We analyzed the \textit{Swift} monitoring data to construct simultaneous X-ray and UV light curves of Mrk~766 (Figure~\ref{fig:swift_lc_ICCF}). The X-ray (0.3--10 keV) light curve exhibits strong variability over timescales of days to weeks, with fractional changes in flux by nearly a factor of two within the month-long monitoring window. In comparison, the UV bands show lower-amplitude variability, yet still demonstrate significant flux changes at the 10--20\% level. The observed differences in variability amplitude are consistent with the expectation that X-ray emission originates in the compact corona, while the UV emission is produced in a more extended region of the accretion disk.
    
    A visual inspection of the light curves suggests correlated trends between the X-ray and UV fluxes. In particular, X-ray flares are often followed by a mild rise in the UVW1 and U-band fluxes, albeit with a reduced amplitude. To assess this connection, we computed the Interpolated Cross-Correlation Function (ICCF; \citealt{Peterson_2004}) with flux randomization/random subset selection (FR/RSS; \citealt{Peterson-FR/RSS-1998-PASP}) to estimate uncertainties on the measured lags using the \texttt{PyCCF} package \citep{PyCCF_2018}\footnote{\texttt{PyCCF} (python based code) used to compute ICCFs is available at \url{https://bitbucket.org/cgrier/python_ccf_code/src/master/}}. We found that X-ray and UV variations are moderately correlated in U and UVW1 bands. No significant correlation is observed between UVM2, UVW2 and X-ray bands. For the U and UVW1 bands the ICCF and FR/RSS centroid distributions peak towards the right which suggest positive lags of $\tau_{\mathrm{U}} = 5.9^{+4.1}_{-5.7}$ hours and $\tau_{\rm W1} = 7.1^{+3.5}_{-5.7}$ hours respectively, with the UV variations trailing the X-rays (Figure \ref{fig:swift_lc_ICCF}). While we are able to measure time lags in these two UV bands, they are associated with large uncertainties. This is due to the fact that the measured lags are smaller than the average sampling rate by a factor $\sim2$ leading to these significant uncertainties. 
    
    The timing behavior observed here is consistent with the reprocessing scenario in which variable X-ray emission irradiates the accretion disk, causing thermal fluctuations in the UV-emitting regions. 
    The measured delay at U-band corresponds to a light-travel distance of $R \sim c\tau \approx 6.36\times10^{14}\ \mathrm{cm}$ or $R \sim 3.4\times10^{3}\,r_{\mathrm{g}}$ for $M_{\bullet}\sim1.26\times10^{6}M_{\odot}$.
    In the framework of a standard geometrically thin optically thick \citep[e.g.][]{shakura1973} accretion disk, the characteristic radius predominantly emitting at wavelength $\lambda$ can be estimated from the temperature profile with the scaling relation $R(\lambda) \propto (M\dot{M})^{1/3} \lambda^{4/3}$. Using a commonly adopted normalization (e.g., \citealt{Cackett2007, Fausnaugh2016, Edelson_2019ApJ}) this can be expressed as
    
    \begin{equation}
    R(\lambda) \approx 10^{15}
    \left(\frac{M}{10^8 M_\odot}\right)^{2/3}
    \left(\frac{\lambda}{5000\,\text{\AA}}\right)^{4/3}
    \left(\frac{L}{L_{\rm Edd}}\right)^{1/3}
    \ \mathrm{cm}.
    \end{equation}

    For $M \sim 10^{6} M_{\odot}$, $L/L_{\rm Edd} \sim 0.1-1$, and $\lambda \sim 3500$~\AA\ (U-band), this yields $R_{\mathrm{U}} \sim 1.6 \times 10^{13-14}$~cm, which is in good agreement with the lag-derived scale. This consistency supports a scenario in which the UV variability arises from reprocessing of coronal X-ray emission in the accretion disk, broadly consistent with the UV-emitting radii of a standard thin accretion disk. The moderate ICCF amplitude ($\sim$0.4--0.6) and the broad centroid distribution may indicate that only part of the UV variability is attributable to X-ray reprocessing, with the remainder likely due to intrinsic disk fluctuations \citep{Gardner_2017}. However, given the poor sampling rate of the current light curves, it is not possible to determine with certainty the cause of these wide centroid distributions and moderate UV/X-ray correlation strengths. In future multi-wavelength observations need to be performed with improved cadence to precisely measure the inter-band time lags. This will help accurately probe the interplay of different physical processes that maybe affecting UV/X-ray time lags in Mrk~766.

\section{Discussion}
\label{sec:discussion}
\subsection{Multi-phase Outflows and Line-of-Sight Obscuration}
\label{subsec:outflows}

Our multi-epoch \rgs\ and \epn\ analysis reveals that the soft X-ray spectrum of Mrk~766 is governed by a stratified, multi-phase ionized outflow. 
We identify three distinct components spanning mildly ionized ($\log \xi \lesssim 1$), moderately ionized ($\log \xi \sim 1.8$--$2$), 
and highly ionized ($\log \xi \gtrsim 2$--$3$) gas. The absorber exhibits variability in column density on timescales of $\sim$1 day, 
indicating a dynamically evolving and inhomogeneous medium. The covering fraction increases systematically by $\sim 10\%$, albeit with significant intrinsic scatter, 
as the continuum flux rises by a factor of $\sim 3.4$, supporting a radiation-driven outflow scenario. The presence of substantial turbulent velocities 
in the highly ionized phase further suggests that this component is associated with an accelerating disk wind rather than a static circumnuclear absorber \citep{Proga_2007, Higgin_2014}.
Across the nine \xmm\ epochs, the observed spectral variability is primarily driven by changes in the partial covering fraction 
(Figure~\ref{fig:EPIC-pn_rgs_anim}), while the intrinsic continuum slope remains comparatively stable (Table~\ref{tab:summary_table}). 
This behavior is consistent with previous studies of Mrk~766 that invoke clumpy BLR-scale clouds transiting the line of sight 
\citep{Risaliti_2011, Buisson_2018}, embedded within a more diffuse ionized medium. 

We note that \citet{Zatarain_2025} recently analysed a time-averaged spectrum combining four of the same 2005 \textit{XMM-Newton}/RGS observations used here, and reported two photoionized warm absorbers at $\log\xi = 2.15 \pm 0.05$ and $\log\xi = -0.58 \pm 0.11$, together with a collisionally ionized absorber (CIA, $T \sim 51$~eV). The somewhat higher ionization of their dominant warm absorber compared to our $\langle\log\xi_2\rangle \sim 0.94$ arises from two key differences. First, their analysis uses a stacked, high-signal-to-noise spectrum and a self-consistent broadband SED from simultaneous \epn\ and OM data to define the ionizing continuum, whereas our \texttt{warmabs} model adopts a fixed SED at table generation time; the shape of the ionizing continuum between 1 and 1000~Ryd directly determines the inferred ionization balance. Second, we do not include a CIA component, which modifies the ionization parameters of the remaining photoionized phases when included. The CIA component reported by \citet{Zatarain_2025} was detected at high significance in the stacked spectrum thanks to the increased signal-to-noise in the 15--20~\AA\ region; in the individual-epoch spectra analysed here, the photon statistics are insufficient to independently constrain such a component, and including it would overparameterize the fits. The \texttt{photemis} ionization 
parameters we recover ($\langle\log\xi_3\rangle \sim 1.763 \pm 0.207$) are in good agreement with the \texttt{pion} emission component of \citet{Zatarain_2025} ($\log\xi = 1.56 \pm 0.05$), supporting a consistent physical picture of the photoionized emission region.

Our results strengthen this interpretation by directly linking 
the warm absorber properties inferred from high-resolution \rgs\ spectroscopy to the partial-covering behavior observed in CCD spectra.
We also find a tentative positive correlation between covering fraction and the 2--10 keV continuum flux, suggesting that the absorber geometry 
evolves with accretion state. Both simulations and observations indicate that higher Eddington ratios drive stronger outflows \citep[e.g.,][]{proga1998, giustini2019}. 
In this scenario, a fixed line of sight intersects a larger effective column of ionized material during brighter states, 
naturally producing the observed luminosity-dependent covering behavior \citep{Tombesi_2013, Matzeu_2017}. Although the statistical significance of this trend remains modest,
its consistency with theoretical expectations and observations of other luminous NLSy1 galaxies supports a physically meaningful connection.

\begin{figure}[htb!]
\centering
\includegraphics[width=0.45\textwidth]{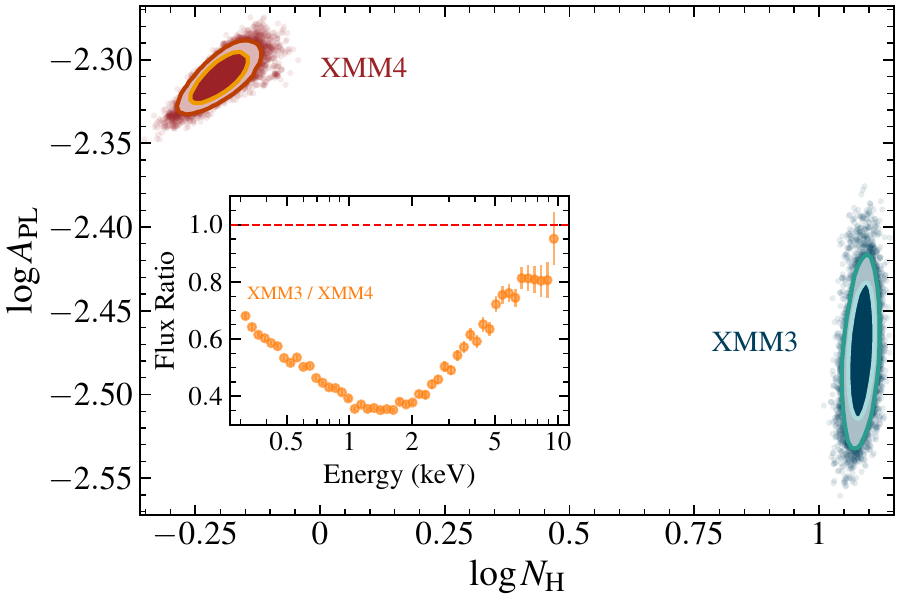}
\caption{Scatter plot of the line-of-sight column density $N_{\rm H, los}$ versus the intrinsic power-law normalization ($\log A_{\rm PL}$) derived from the \texttt{zxipcf} fits to the EPIC-pn spectra. The two distinct clusters correspond to the XMM3 and XMM4 epochs, separated by $\Delta t \simeq 2$~days, and are clearly offset in $N_{\rm H, los}$, indicating rapid variability of the absorbing column. The inset shows the energy-dependent flux ratio (XMM3/XMM4), which exhibits a pronounced suppression at soft X-rays, consistent with enhanced absorption during the XMM3 epoch rather than intrinsic continuum variations.}
\label{fig:chlook}
\end{figure}

    Figure \ref{fig:chlook} shows the scatter distribution of the line-of-sight column density $N_{\rm  H, los}$ intrinsic power-law normalization derived from the \texttt{zxipcf} fits to the EPIC-pn spectra for two closely separated epochs ($\Delta t \simeq 2$~days). The two clusters are clearly separated in $N_{\rm  H, los}$, indicating that the absorber undergoes a substantial reconfiguration within $\Delta t$.
    The inset in Figure~\ref{fig:chlook} shows the energy-dependent flux ratio (XMM3/XMM4), which exhibits a suppression at soft X-rays and recovers at higher energies. This spectral curvature is characteristic of variable absorption, suggesting that the observed variability is primarily driven by changes in the line-of-sight column density rather than intrinsic continuum variations.
    Assuming that the observed variability timescale traces the dynamical timescale, the radius associated with the dynamical timescale ($t_{\rm dyn} = \sqrt{R^3/GM_{\rm BH}}$) for this $N_{\rm H}$ variation, the inferred characteristic radius of the absorber is corresponds to 900~R$_{\rm g}$, adopting a black hole mass of $1.26 \times 10^6 M_{\rm \odot}$. While this estimate is subject to uncertainties in the physical origin of the variability (e.g., cloud transit versus ionization response), it provides a characteristic scale for the absorber location.
    Such rapid variability strongly suggests that the absorbing material is located at sub-parsec scales, likely within or comparable to the broad-line region or the inner torus. This interpretation of spatial scale is consistent with previous studies of rapidly variable X-ray absorbers in Seyfert galaxies and NLSy1s, which associate column density variability with clumpy circumnuclear material on BLR scales \citep{Risaliti_2002, Risaliti_2011, Markowitz_2014}. 

    While both Keplerian cloud motion and radiation-driven outflows are viable at these radii, the observed increase in covering fraction with continuum flux favors a radiation-pressure-regulated wind scenario \citep[e.g.,][]{Proga_2000, Proga_2007, Giustini_2019}. In this framework, enhanced ionizing flux can lift denser clumps into the line of sight, naturally producing the observed luminosity-dependent obscuration. We therefore interpret the absorber as a stratified, multi-phase clumpy wind launched from the inner accretion disk, whose transverse motion drives rapid $N_{\rm H, los}$ variability, while its vertical response to radiation pressure modulates the covering fraction.
    
    In contrast, purely Keplerian motion would not naturally predict such a flux-dependent covering behavior. We therefore interpret the observed behavior in terms of a clumpy, partially ionized absorber, in which individual clouds intermittently intersect the line of sight to the X-ray emitting corona, producing the observed variability in $N_{\rm H, los}$ and apparent flux changes. We therefore interpret the absorber as a stratified, multi-phase clumpy wind launched from the inner disk region, whose transverse motion produces rapid column density variability, while its vertical response to radiation pressure modulates the covering fraction. The observed behavior is consistent with a disk-wind origin rather than a static circumnuclear obscurer.

\subsection[Origin and Evolution of the Fe K alpha Emission]{Origin and Evolution of the \fekalpha\ Emission}
\label{fekalpha_origin_evoln}
    The \xrism-Resolve high-resolution spectroscopy shows that the \fekalpha\ line profile is composite, consisting of a stable, narrow 6.4~keV core with an additional broadened component that cannot be reproduced alone by a narrow emission line originating from a distant torus \citep[e.g.][]{Murphy_Yaqoob_2009}. The inferred inner radius of the broadened component is $R_{\rm{in}}\sim 40-60~r_{\rm{g}}$, requiring relativistic broadening. Recent \xrism\ observations of bright Seyfert galaxies, such as NGC~4151, NGC~3783, and Mrk~279, have revealed that the neutral Fe K$\alpha$ emission often consists of multiple distinct components arising from different spatial scales: the molecular torus at radii $r \gtrsim 10^{4}~r_g$, an X-ray-emitting extension of the broad-line region (BLR) at $r \sim 10^{3}$--$10^{4}~r_g$, and the inner accretion disk at $r \lesssim 100~r_g$ (e.g., \citealt{xrism_ngc4151, Mehdipour_ngc3783_2025, J_Miller_mrk279_2025}). The long-term \epn\ monitoring reveals a coherent picture in which the broad Fe K$\alpha$ flux closely tracks the continuum variations, while the narrow core remains comparatively stable across epochs (see Figure~\ref{fig:Feline_params_epochs} and panel \textbf{(c)} in Figure~\ref{fig:correlation_plot}). The rapid response of the broad component implies a short light-crossing timescale and therefore a compact emitting region, consistent with an origin in the inner accretion disk ($r \lesssim 10$--$100~r_g$). In contrast, the lack of significant variability in the narrow core indicates reprocessing in distant material on parsec scales, such as the molecular torus or outer BLR. 

The unprecedented energy resolution of the \textsc{XRISM}/Resolve 
spectrum allows us to place a direct constraint on the velocity 
width of the narrow Fe~K$\alpha$ core, providing an independent 
estimate of the location of the distant emitter. Freeing the 
\texttt{gsmooth} broadening parameter $\sigma_{\rm 6keV}$ in 
the spectral fit yields a 90\% upper limit of 
$\sigma_{\rm 6keV} < 0.0757$~keV, corresponding to a velocity 
dispersion upper limit of $\sigma_v < 3546$~km~s$^{-1}$ 
(FWHM~$< 8351$~km~s$^{-1}$). Applying the virial relation:
\begin{equation}
    R \gtrsim \frac{GM_{\bullet}}{\sigma_v^2} 
    \approx 7.2 \times 10^{3}\ r_{\rm g},
\end{equation}
adopting $M_{\bullet} = 1.26 \times 10^{6}\ M_{\odot}$, this places the narrow Fe~K$\alpha$ emitter 
at $r \gtrsim 7\times 10^{3}~r_{\rm g}$, consistent with an 
origin in the outer BLR at $r \sim 10^{3}$--$10^{4}~r_{\rm g}$ 
or the inner wall of the molecular torus at 
$r \sim 10^{4}$--$10^{5}~r_{\rm g}$. This is wholly 
inconsistent with an origin in the innermost accretion disk 
($r \lesssim 100~r_{\rm g}$), where the relativistically 
broadened emission is independently accounted for by the 
\texttt{diskline} component. The key \texttt{diskline} 
parameters remain unaffected by freeing $\sigma_{\rm 6keV}$, 
with the inferred inner disk radius 
$\log R_{\rm in} = 1.61^{+0.96}_{-0.27}~r_{\rm g}$ fully 
consistent with the baseline value of 
$\log R_{\rm in} = 1.63^{+0.50}_{-0.24}~r_{\rm g}$, 
confirming that the broad and narrow Fe~K$\alpha$ components 
trace physically and spatially distinct emission regions.
    
    These results are in excellent agreement with recent \xrism\ findings, providing complementary time-domain evidence for a radially stratified origin of the \fekalpha\ emission, where the broad component traces the inner accretion disk and the narrow core arises from more distant reprocessing material. In particular, \xrism\ observations of MCG–6-30-15 \citep{Brenneman_2025} likewise require a strong relativistically broadened \fekalpha\ component to adequately describe the data. Together with our results for Mrk~766, this points toward a scenario in which variations in coronal illumination may play a key role in regulating the spatial origin and strength of relativistic reflection in Seyfert galaxies.

\subsection[Linking Fe K alpha Reflection and UV--X-ray Reprocessing]{Linking \fekalpha\ Reflection and UV--X-ray Reprocessing}
\label{feklapha_UV-Xray}
    The UV--X-ray lag, the \fekalpha\ line morphology, and the line-of-sight covering-fraction behavior can all be interpreted within a single accretion--corona--disk-wind geometry. The quasi-simultaneous \xrism\ and \swift\ observation of Mrk~766 reveals the connection between the UV--X-ray time lag and \fekalpha\ complex origin morphology. The spectral modeling from \xrism-Resolve \& Xtend data places the inner-radii of the broad \fekalpha\ emission at $R_{\rm{in}}\sim 40-60~r_{\rm{g}}$, indicating that the inner accretion disk is directly illuminated by the compact X-ray corona. This same coronal variability drives thermal reprocessing in the outer disk, producing the observed UV delay of $\sim 5$ hours \citep{GuilbertRees1988, Cackett2007}. The lag corresponds to radii of $\sim 10^{15}~\rm{cm}$ (a few times $10^3~r_{\rm{g}}$), consistent with UV-band emission from a standard thin disk around a $\sim 10^6 M_\odot$ black hole \citep{McHardy2014, Fausnaugh2016}. Thus, the broad \fekalpha\ line traces the innermost response to coronal fluctuations, while the UV lag reflects the outer-disk response to the same driving X-ray emission, but on light-travel timescales orders of magnitude larger.

It is important to clarify the physical geometry implied by 
the simultaneous presence of a compact, rapidly varying corona 
and a broad Fe~K$\alpha$ emitter with inner radius $R_{\rm in} 
\sim 40$--$60~r_{\rm g}$. The inferred $R_{\rm in}$ represents 
the innermost radius of the disk contributing 
significantly to the observed Fe~K$\alpha$ fluorescence, and 
should not be interpreted as the spatial extent of the corona. 
A compact corona confined within a few tens of $r_{\rm g}$ 
naturally illuminates a disk beginning at this radius, with 
the interior region likely occupied by a hot, optically thin 
plasma contributing negligibly to neutral Fe~K$\alpha$ 
fluorescence. Given the high Eddington ratio of Mrk~766, 
a lamppost-like coronal geometry \citep{MiniuttiFabian2004} 
illuminating a moderately truncated disk is the most 
physically natural interpretation. The rapid variability 
of the broad Fe~K$\alpha$ flux on timescales of 
$\sim 10^{4}$~s independently constrains the coronal size 
to $r_{\rm corona} \lesssim c\,\delta t \approx 250~r_{\rm g}$, 
confirming that the corona remains compact and well within 
the truncation radius, with no physical inconsistency between 
the two.

    As the X-ray luminosity increases, radiation pressure is expected to inflate the disk wind, increasing its vertical extent and the likelihood that our line of sight intersects a larger column of ionized gas. This results in a higher effective covering fraction, as measured by the \texttt{zxipcf} component in brighter states (Figure~\ref{fig:correlation_plot}), and modulates the observed line-of-sight absorption. At the same time, the close tracking between the broad \fekalpha\ flux and the 2–10~keV continuum (Figure~\ref{fig:Feline_params_epochs}) indicates that the inner accretion disk responds promptly to variations in the coronal emission, likely due to a more centrally concentrated illumination pattern driven by changes in the coronal geometry. In contrast, the UV emission exhibits a delayed response, consistent with reprocessing in the outer disk on light-travel timescales of several hours.

    Taken together, the \fekalpha\ morphology, absorber variability, and UV--X-ray lag can be interpreted within a unified disk--corona--wind framework. A compact X-ray corona illuminates the inner disk, producing the rapidly varying broad \fekalpha\ component, while simultaneously driving thermal reprocessing in the outer disk that gives rise to delayed UV variability. As the accretion rate increases, radiation pressure inflates the disk wind vertically, enhancing the line-of-sight covering fraction and modulating how efficiently coronal photons illuminate different disk radii \citep{MiniuttiFabian2004}. This geometry naturally explains why the broad \fekalpha\ flux responds promptly to continuum changes, why the narrow core remains stable, and why the UV emission follows the X-rays with a measurable delay. The moderate ICCF amplitudes further suggest that only part of the UV variability is driven by reprocessing, with intrinsic disk fluctuations contributing as well \citep{Edelson2015}.

\section{Conclusions}
\label{sec:conclusions}
    The results we present in this study clearly demonstrate what \xrism\ adds beyond \xmm\ observations for this NLSy1 source Mrk~766, in which high-resolution X-ray spectroscopy connects accretion, reflection, and ionized outflows to disentangle the true structure of the \fekalpha\ complex into several kinematics components. The combination of long-term \xmm\ monitoring with a \xrism\ observation demonstrates that the origin and morphology of the \fekalpha\ emission complex in this source cannot be described by a single, static reprocessing region, but instead reflects a dynamic environment in which the inner-disk, distant material, and line-of-sight absorbers all contribute in a luminosity-dependent manner. We used a quasi-phenomenological framework that includes both reprocessing in distant material and a relativistically broadened \texttt{diskline} component, enabling us to disentangle the narrow and broad contributions to the \fekalpha\ emission line and morphology of the line-emitting region which places the inner radius of the region responsible for the broadened emission at approximately $R_{in} \simeq 40-60~r_{\rm g}$. A multi-epoch investigation of the Fe K$\alpha$ complex based on \textit{XMM-Newton} data shows that the broad line component responds sensitively to changes in the continuum flux, closely following its variability pattern. In contrast, the narrow feature remains largely invariant, consistent with an origin in distant material such as the torus.

\begin{acknowledgments}

    We thank the anonymous referee for useful comments and suggestions that helped to improve the manuscript. This research has made use of data and/or software provided by the High Energy Astrophysics Science Archive Research Center (HEASARC), which is a service of the Astrophysics Science Division at NASA/GSFC. This work made use of data supplied by the UK Swift Science Data Centre at the University of Leicester. We thank Gulab Dewangan for useful comments that helped to improve the manuscript. A.S. and R.C. thank IUCAA for their hospitality and usage of their facilities during their stay at different times as part of the university associateship program, ISRO for support under the \textit{AstroSat} archival data utilization program and ANRF for a SURE grant (File No. SUR/2022/001503). R. C. thanks ANRF for an ARG grant (ANRF/ARG/2025/003315/PS).
    
\end{acknowledgments}

\section*{Data availability} \label{sec:data_avl}

    This study makes use of data publicly available in the HEASARC archive (\url{https://heasarc.gsfc.nasa.gov/db-perl/W3Browse/w3browse.pl}) maintained by HEASARC. Reduced and analyzed data products not hosted in the archive may be provided by the corresponding author upon reasonable request.

\facilities{XMM-Newton, Swift(XRT and UVOT), XRISM}

\software{ Astropy \citep{AstroPy_2013, AstroPy_2018}, Emcee \citep{emcee_2013}, Xspec \citep{Arnaud_Xspec,Dorman_Xspec,Dorman_2003}, HEAsoft \citep{heasoft_2014}, Matplotlib \citep{Hunter:2007}
          }

\appendix
\restartappendixnumbering

\section{Non-X-ray background in the Resolve spectrum of Mrk~766}
\label{app:nxb}
\begin{figure}[h]
\centering
\includegraphics[width=\columnwidth]{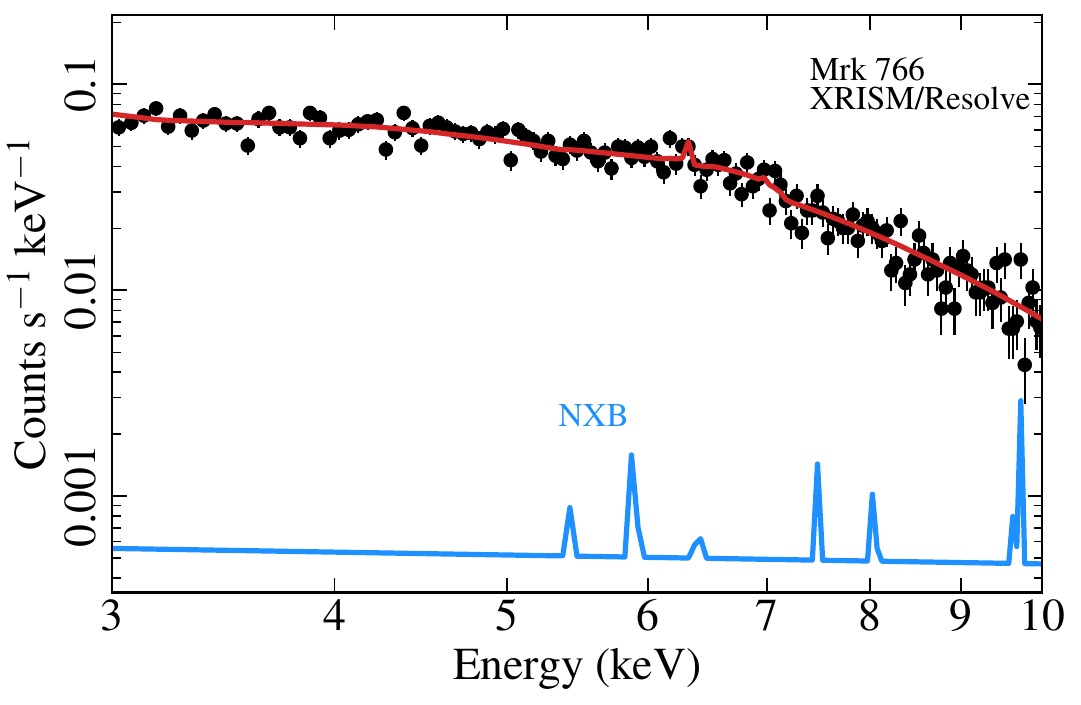}
\caption{Comparison of the count rate spectra of the source (black) and the NXB (blue) in the XRISM/Resolve observation of Mrk 766 in 3--10 keV band. The best-fit model, described in Section \ref{subsec:xrism_spectra}, is shown in red. The NXB contribution across the bandpass is mostly minimal.}
\label{fig:nxb}
\end{figure}

As shown in Figure~\ref{fig:nxb}, the Non-X-ray background (NXB) contribution across the 3--10 keV band is substantially lower than the source count rate. Although several weak instrumental features are present in the NXB spectrum, their amplitudes are more than an order of magnitude below the source counts in the Fe-K band (6--7 keV). Therefore, the broadened \fekalpha\ emission profile is not significantly affected by the NXB contribution. Likewise, the NXB level is too small to produce any appreciable change in the broadband continuum slope. The details methodology adopted for the Resolve data reduction and NXB modeling is described in \href{https://heasarc.gsfc.nasa.gov/docs/xrism/analysis/nxb/nxb_spectral_models.html}{NXB Spectral Models} and \href{https://heasarc.gsfc.nasa.gov/docs/xrism/analysis/nxb/resolve_nxb_db.html}{Resolve NXB Database and Spectral Extraction Recipes} as provided by the XRISM Science team.

\section{Sensitivity of the Inferred Inner Disk Radius to Assumed Inclination}
\label{app:rin_incl}

\begin{figure}
    \centering
    \includegraphics[width=\columnwidth]{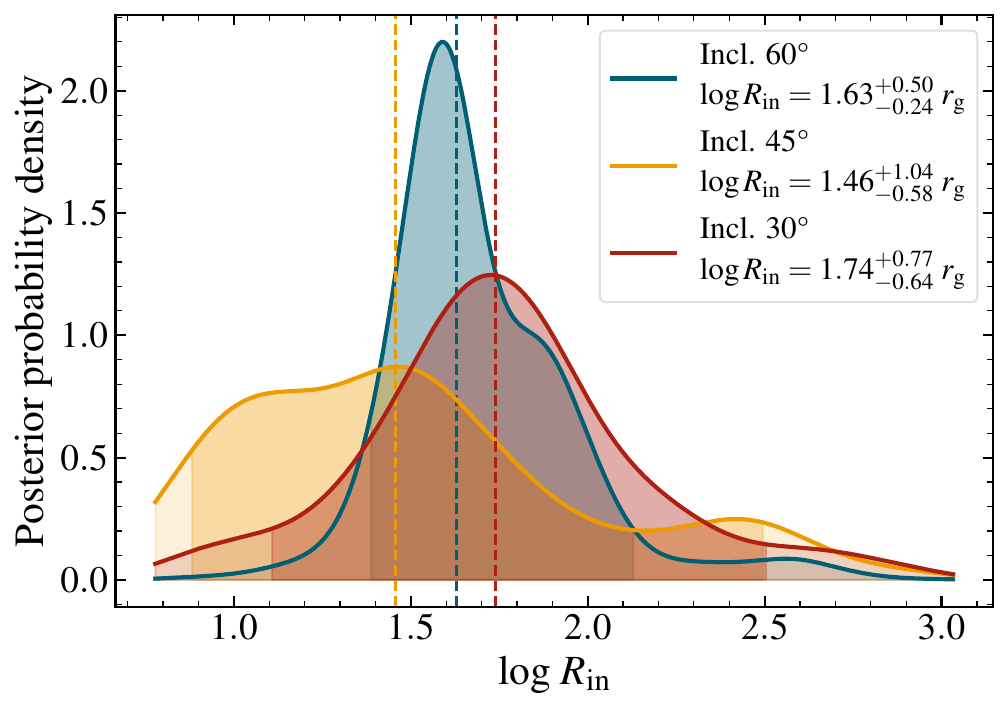}
    \caption{
        Marginalized posterior distributions of $\log R_{\rm in}$ 
        derived from MCMC fitting of the \texttt{diskline} model to 
        the \textit{XRISM}/Resolve Fe~K$\alpha$ spectrum, for three 
        assumed disk inclinations 
        ($i = 30^{\circ}$, $45^{\circ}$, and $60^{\circ}$). 
        Shaded regions indicate the 90\% credible intervals 
        (5th--95th percentile). Vertical lines mark the median values. 
        The distributions overlap substantially, indicating that the 
        inferred $R_{\rm in}$ is not strongly sensitive to the assumed 
        disk inclination.
    }
    \label{fig:rin_incl}
\end{figure}

To assess the robustness of the inner disk radius $R_{\rm in}$ derived from the \texttt{diskline} model, we repeated the \textit{XRISM}/Resolve spectral fit with the disk inclination fixed at three values: $i = 30^{\circ}$, $45^{\circ}$, and $60^{\circ}$, while keeping all other parameters at their baseline values. For the \texttt{diskline} component, the second model parameter,\footnote{\url{https://heasarc.gsfc.nasa.gov/docs/software/xspec/manual/node199.html}} \texttt{Betor10}, was fixed at 10 in all cases. In XSPEC, setting \texttt{Betor10} to 10 or greater invokes the standard accretion-disk \citep{shakura1973} emissivity prescription $\epsilon(r) \propto (1-\sqrt{6/r})/r^3$,
which approaches an $r^{-3}$ emissivity profile at large radii (see \citealt{Fabian_1989}). Figure~\ref{fig:rin_incl} shows the marginalized posterior distributions of $\log R_{\rm in}$ for the three assumed inclinations, derived from the MCMC chains. The best-fit inner radii are $\log R_{\rm in} = 1.63^{+0.50}_{-0.24}$, $1.46^{+1.04}_{-0.58}$, and $1.74^{+0.77}_{-0.64}~r_{\rm g}$ for $i = 60^{\circ}$, $45^{\circ}$, and $30^{\circ}$, respectively, where the uncertainties represent the 90\% credible intervals. Although the posterior peaks shift slightly with the assumed inclination, the distributions overlap substantially within their 90\% credible intervals. The inferred inner radius remains within the inner few tens of gravitational radii in all cases, demonstrating that our primary conclusion that the broad \fekalpha\ emission originates from the innermost accretion disk is robust against reasonable variations in the assumed disk inclination.

\bibliography{mrk766}{}
\bibliographystyle{aasjournalv7}

\end{document}